\documentclass[aps, prd, twocolumn, superscriptaddress, showpacs, floatfix]{revtex4}

\usepackage[dvips]{graphicx}
\usepackage{bm,amsfonts, mathtools}
\usepackage{color}
\newcommand{\w}[1]{\bm{#1}}

\newcommand{\der}[2]{\frac{\partial #1}{\partial #2}}

\newcommand{\dder}[2]{\frac{\partial^2 #1}{\partial #2 ^2}}
\newcommand{\be}{\begin{equation}}
\newcommand{\ee}{\end{equation}}
\newcommand{\bea}{\begin{eqnarray}}
\newcommand{\eea}{\end{eqnarray}}

\let\ph\varphi
\let\th\theta

\newcommand{\lapang}{\Delta_{\theta\varphi}}
\let\ph\varphi
\let\th\theta
\newcommand{\mw}{\mathcal{W}}
\newcommand{\mx}{\mathcal{X}}

\begin{document}
 
\title{Excised black hole spacetimes: quasi-local horizon formalism applied to
  the Kerr example}

\author{Nicolas Vasset} 
\email{Nicolas.Vasset@obspm.fr}

\author{J\'{e}r\^{o}me Novak}
\email{Jerome.Novak@obspm.fr}
\affiliation{Laboratoire Univers et Th\'eories, Observatoire de
  Paris, CNRS, Universit\'e Paris Diderot, 5 place Jules Janssen,
  F-92190 Meudon, France}

\author{Jos\'e Luis Jaramillo}
\email{jarama@iaa.es}
\affiliation{Instituto de Astrof\'{\i}sica de Andaluc\'{\i}a, CSIC,
  Apartado Postal 3004, E-18080 Granada, Spain}
\affiliation{Laboratoire Univers et Th\'eories, Observatoire de
  Paris, CNRS, Universit\'e Paris Diderot, 5 place Jules Janssen,
  F-92190 Meudon, France}
\date{March 31, 2009}
 
\begin{abstract}
  We present a numerical work aiming at the computation of excised initial
  data for black hole spacetimes in full general relativity, using Dirac gauge
  in the context of a constrained formalism for the Einstein equations.
  Introducing the isolated horizon formalism for black hole excision, we
  especially solve the conformal metric part of the equations, and assess the
  boundary condition problem for it. In the stationary single black
  hole case, we present and justify a no-boundary treatment on the black hole
  horizon. We compare the data obtained with the well-known analytic Kerr
  solution in Kerr-Schild coordinates, and assess the widely used conformally
  flat approximation for simulating axisymmetric black hole spacetimes. Our
  method shows good concordance on physical and geometrical issues, with the
  particular application of the isolated horizon multipolar analysis to
  confirm that the solution obtained is indeed the Kerr spacetime. Finally, we
  discuss a previous suggestion in the literature for the boundary conditions
  for the conformal geometry on the horizon.
\end{abstract}

\pacs{
04.25.Dg,    %Numerical studies of black holes and black-hole binaries
04.20.Ex,    %Initial value problem, existence and uniqueness of solutions
04.70.-s,    %Physics of black holes
04.25.D-     %Numerical relativity 
}

\maketitle

%%%%%%%%%%%%%%%%%%%%%%%%%%%%%%%%%%%%%%%%%%%%%%%%%%%%% 
\section{\bf Introduction} 
\label{sec:intro}
%%%%%%%%%%%%%%%%%%%%%%%%%%%%%%%%%%%%%%%%%%%%%%%%%%%%% 

Trying to accurately describe black holes solutions as evolving physical
objects in numerical simulations is of direct interest in
astrophysics. Numerical simulations have made a great leap forward in the past
few years, mainly with the first stable simulations of black hole mergers in
full general relativity by Pretorius \cite{Pret05}, Campanelli et
al. \cite{Camp}, Baker et al. \cite{BC06}, and a few other groups (see
\cite{Pret07} for a review). Several of these simulations model black holes in their equations by punctures. These punctures basically
change the topology of spacetime to handle evolution of singular objects (see
\cite{BrBr} for the first proposal of this method).

Notable exceptions to this are references
\cite{Pret05,Caltech,SziPolRez07,Spe07}, that use an excision approach for black hole
evolution: the simulations only evolve the spacetime outside two-spheres that
are supposed to encircle the black hole singularities.

In the same spirit, we are here trying to describe black holes as physical
objects characterized by their horizons. Defining the physical laws for event
horizons of black holes has been notably done by \cite{HawHar}, \cite{Dam79}
and \cite{TPM86} in a so-called {\it membrane paradigm}. The aim was to describe them as fluid-like
two-membranes with physical properties. However, applying evolution laws to
event horizons is problematic due to their teleological (non causal)
behavior (see for example \cite{Boo05} for a description of the
phenomenon). Being defined as a global property of spacetime, a local
notion of causality actually does not apply to event horizons.

Alternative local characterizations have been formulated in the past 15 years
by \cite{Hay94}, \cite{AshBF} and \cite{AshKr03}. They are based on the
concept of trapped surfaces, dating back to Penrose's singularity
theorem \cite{Pen65}. Defined locally, those objects behave in a causal way
in a general dynamical context, with local evolution laws following from the
projection of Einstein equations, e.g the Navier-Stokes \cite{Gourg05} ``fluid bubble''
analogy or the area evolution law in \cite{GouJar06b}. For this work we shall
use local characterizations for isolated horizons, prescribing the physics of
non-evolving black hole horizons.

Following the prescriptions of \cite{Cook02}, \cite{GJM04}, \cite{GJ06} and
pursuing the numerical explorations of \cite{CookPf04,CauCooGri06} and \cite{JAL07} among
others, we try to numerically implement those objects as boundary conditions
imposed on the (3+1) form of Einstein equations, in a three-slice excised by a
two-surface (single black hole case). This is done here using the fully
constrained formalism (FCF) of \cite{Bon04} (see also \cite{Cord1}), with maximal slicing and Dirac
gauge, based on the (3+1) formalism, and with spectral methods-based numerical
resolution using the LORENE library \cite{LORENE}. An important point here is
that we drop out the usual conformal flatness hypothesis and solve for the
conformal geometry, so that we can exactly recover a slice of a stationary rotating
vacuum spacetime.

Contrary to free evolution schemes, which are the most used prescription for
(3+1) simulations in numerical relativity, an important feature of constrained
schemes is the necessity to solve constraints on each three-slice, in the form of
elliptic equations. These equations generally need additional conditions to
be imposed on grid boundaries, following reasonable geometrical and physical
prescriptions. Our approach particularly requires a specific handling of
boundary conditions for the two dynamical gravitational degrees of freedom. This
is a crucial point of our calculation; we will justify and apply here a
no-boundary treatment for these quantities.

The paper is organized as follows: we first review in Sec.~\ref{s:isol_hor} fundamental geometrical
properties associated with isolated horizons in general relativity. In Sec. ~\ref{s:FCF}, we quickly give the basic
features of the fully constrained formalism and the methods we use to treat
the conformal part. Sec.~\ref{s:BCs} discusses implementation of boundary
conditions for the system of equations, and specifically discusses the
conformal metric part. Sec.~\ref{s:resus} gives the numerical results
obtained, and confronts them to a battery of tests characterizing the physics
of the solutions. A discussion follows in Sec.~\ref{s:discus}, in
regard of previous works concerning the computation of the conformal part in
the black hole initial data problem. We also raise the question of
applicability of this scheme to other more general astrophysical cases.

Throughout all this paper, Greek letters will denote indices spanning from 0
to 3, Latin indices from $k$ to $m$ shall denote indices from $\{1,2,3\}$, and
indices from $a$ to $c$ have the range of $\{2,3\}$. All formulae and
values are given in geometrical units ($G=c=1$). We also use the Einstein
summation convention.
 
%%%%%%%%%%%%%%%%%%%%%%%%%%%%%%%%%%%%%%%%%%%%%%%%%%%%%%%%%%%%%%%%%%%%%%%%%%%%%%%%% 
\section{Isolated horizons as a local description of black hole
  regions}\label{s:isol_hor}
%%%%%%%%%%%%%%%%%%%%%%%%%%%%%%%%%%%%%%%%%%%%%%%%%%%%%%%%%%%%%%%%%%%%%%%%%%%%%%%%%

\subsection{Trapped surfaces and expansion}

The concept of a trapped surface in a Lorentzian frame has been
first defined by Penrose in 1965 \cite{Pen65} in connection with the
singularity theorems. It relies on the notion of expansion of the
light rays emitted from a surface, that
we explain here. We start by a closed spacelike two-surface $\mathcal{S}$
embedded in spacetime, topologically related to a two-sphere. We assign to it a
two-metric $q_{ab}$ induced by the ambient four-metric $g_{\mu\nu}$, and its
associated area form $\epsilon^{\mathcal{S}}_{ab}$. Two future null
directions orthogonal to $\mathcal{S}$ are associated with this surface. Representative vector
fields are denoted $\ell^\mu$ and $k^\mu$, being respectively oriented
outwards and inwards. We can for example assume that our spacetime is
asymptotically flat, so the orientation can be defined without ambiguity.

The expansion $\theta_{\ell}$ of $\mathcal{S}$ along $\ell^\mu$ is the area rate
of change along this vector:
$\mathcal{L}_{\ell}\,{\epsilon}^{\mathcal{S}}_{ab} =
\theta_{\ell}\,\epsilon^{\mathcal{S}}_{ab}$. $\mathcal{L}$ is the Lie derivative, here along the vector $\ell^\mu$. Same definition goes for the
vector field $k^\mu$. For a two-sphere embedded in a Minkowski
spacetime (flat metric), we have typically $\theta_{k}
< 0$ and $\theta_{\ell} > 0$. $\mathcal{S}$ is said to be a trapped
surface if both expansions are negative or zero: $\theta_{\ell} \leq 0$ , $
\theta_{k} \leq 0$. This clearly characterizes strong local curvature. A
marginally (outer) trapped surface will be characterized by $\theta_{\ell} = 0$
and $ \theta_{k} \leq 0$. Two theorems make the connection between those
objects and black holes: provided the weak energy condition holds, the
singularity theorem of Penrose \cite{Pen65} ensures that a spacetime
containing a trapped surface necessarily contains a singularity in its
future. Following this result, provided the cosmic censorship holds, another result
by Hawking and Ellis \cite{HawEll73} conveys that a spacetime containing a
trapped surface necessarily contains a black hole region enclosing this
surface.

Marginally outer trapped surfaces (MOTS)
%, being the outermost trapped surfaces of a trapped region, 
are intended as models for the black hole boundary (see \cite{Senov08}
for a discussion of its relation with the boundary of the black hole trapped region). Local horizons (trapping horizons for \cite{Hay94}, isolated and
dynamical horizons in \cite{AshBF}, \cite{AshKr03}) are defined as
three-dimensional tubes ``sliced'' by MOTS, with additional geometrical
properties. The isolated horizon case is detailed below; for a review, see
\cite{AshKr04}.
 
In vacuum stationary spacetimes, all horizons are at the same location, which
is also the location of the event and apparent horizons (constructed with outermost MOTSs). 
In the more general
case, and assuming cosmic censorship, local horizons are always situated inside the event horizon in general
relativity.

\subsection{Isolated horizons}

The notion of isolated horizon is aimed at describing stationary black
holes. It is based on the notion of non-expanding horizons, and defined as a three-dimensional tube $ \mathcal{H}$
foliated by MOTS, and with a null vector field $\ell^\mu$ as generator. The
three-metric induced on the tube has then a signature $(0, +, +)$.

We also define a (3+1) spatial slicing for our spacetime, and $\mathcal{S}$ a
two-slice of our isolated horizon at a certain value of the time parameter
$t$. The spacelike two-metric on $\mathcal{S}$ is denoted $q_{ab}$.
  
The shear tensor $\sigma_{ab}$ on the two-surface is defined along $\ell^\mu$ as
 \begin{equation}
  \sigma_{ab} = \frac{1}{2}\left[\mathcal{L}_{\ell}\,q_{ab} - \theta_\ell
    q_{ab}\right].
  \label{e:shear}
\end{equation}

 Using
 the fact that $\theta_{\ell} = 0 $ and the dominant energy condition, the
 Raychaudhuri equation for null tubes \cite{GJ06} ensures that both the shear
 along $\ell^\mu$ and the energy-momentum tensor projected on $\ell^\mu$, and
 evaluated on the surface must vanish: $\sigma_{ab} =0$ and $T_{ab}\ell^{a}
 \ell^{b} = 0$. These are additional properties constraining the geometry of
 the horizon.

 An isolated horizon is also required to be such that the extrinsic geometry of the tube
 is not evolving along the null generators: $[\mathcal{L}_{(\ell)}, D_i^{\cal
   H}] = 0$, where $D_i^{\cal H}$ is the connection on the tube induced
from the ambient spacetime connection.
If this last condition is dropped, we only retrieve a non-expanding horizon (linked to the notion of ``perfect horizon''
 \cite{Haj73}).

 The isolated horizon formalism has already been studied extensively in
 numerics, as a diagnosis for simulations involving black holes, where
 marginally trapped surfaces are found {\it a posteriori\/} with numerical tools
 called apparent horizon finders (See for example~\cite{Thorn},
 \cite{LinNov07}, \cite{AHfinder1}, \cite{AHfinder2}). Let us note that very often,
 apparent horizon finders actually 
 locate MOTS on the three-slice considered (not necessarily outermost ones). A thorough study of geometrical
 properties of isolated horizons located {\it a posteriori\/} can be found in
 \cite{DrKrSch03}.

 In the present paper we employ isolated horizons as an {\it a priori} ingredient in the numerical
 construction of Cauchy initial data for black hole spacetimes.
 More precisely, we impose conditions on the excised surface characterizing it as the slice of a
 non-expanding horizon (see below). This approach to the modeling a black hole horizon in instantaneous
 equilibrium  has been investigated in 
 \cite{Cook02,CookPf04,CauCooGri06,GJM04,DaiJarKri05,GJ06,JAL07}, where a prescription
 for the conformal metric is assumed.
 The main feature of this work is the inclusion of the conformal metric 
 in the discussion, 
 not through analytical prescriptions, but indeed by numerical calculation. This problem  
 has also been recently addressed in  \cite{CookBaum08}.
 In  Sec.~\ref{s:BCs} we tackle the description of our special
 treatment of the conformal metric on the excised boundary and compare with previous
 results, in particular through the numerical recovery of excised Kerr initial
 data.
 
 %Several numerical works using this {\it modus operandi\/} to simulate
 %single, or even black-hole binary initial data include the works of
 %\cite{CookPf04,CauCooGri06} and \cite{JAL07}, and the black-hole binary simulations of
 %\cite{Caltech}. While a comparison with these results will be done thoroughly
% later in this paper,

%%%%%%%%%%%%%%%%%%%%%%%%%%%%%%%%%%%%%%%%%%%%%%%%%%%%%%%%%%%%%%%%%%%%%%%%%%%%%%%%%
\section {A fully constrained formalism for Einstein equations}\label{s:FCF}
%%%%%%%%%%%%%%%%%%%%%%%%%%%%%%%%%%%%%%%%%%%%%%%%%%%%%%%%%%%%%%%%%%%%%%%%%%%%%%%%%
  
All the following is a summary of the physical and technical assumptions set in
\cite{Bon04}. For a review of (3+1) formalism in numerical relativity, the
reader is referred to \cite{York79}, or more recent reviews like
\cite{Gourg07} and \cite{Alc08}.

\subsection{Notations and (3+1) decomposition}
We consider an asymptotically flat, globally hyperbolic four-dimensional manifold
$\mathcal{M}$, associated with a metric $g_{\mu \nu}$ of Lorentzian
signature $(-,+,+,+)$. We define on $\mathcal{M}$ a slicing by spacelike
hypersurfaces $\Sigma_{t}$, labeled by a timelike scalar field $t$; in this
way, the four-metric can be written in its usual (3+1) form:
\begin{equation}
g_{\mu \nu}dx^{\mu}dx^{\nu} = -N^{2}dt^{2} + \gamma_{ij}(dx^{i} +
\beta^{i}dt)(dx^{j} + \beta^{j}dt), 
\label{e:metric}
\end{equation} 
here $N$ and $\beta^{i}$ are the usual lapse scalar field and shift vector
field. $\gamma^{ij}$ is the spacelike three-metric induced on $\Sigma_{t}$.

We also define the second fundamental form of $\Sigma_{t}$, or extrinsic
curvature tensor, as:
 \begin{equation}
K_{\mu \nu}=- \frac{1}{2}{\mathcal{L}_{n}} \gamma_{\mu \nu},
\label{e:Kij}
\end{equation}
with $n^{\mu}$ the future-directed vector field normal to $\Sigma_{t}$. Writing the vacuum Einstein equations with this
formalism, one comes up with the classical (3+1) vacuum Einstein equations system(see
for example \cite{ADM62}):
\begin{eqnarray}
\label{e:3+1} 
R + K^{2} - K_{ij}K^{ij} &=& 0,\\
D_{j} K_{i}^{j} - D_{i}K &=& 0,\\
\frac{\partial}{\partial t} K_{ij} - {\mathcal{L}}_{\beta}K_{ij} &=&
\nonumber\\
-D_{i}D_{j}N &+& N \left\{R_{ij} - 2K_{ik}K^{k}_{j} + KK_{ij} \right\},
\end{eqnarray}
$ D_{i}$ and $R_{ij}$ being, respectively, the connection and the Ricci tensor
associated with the three-metric $\gamma_{ij}$. Quantities without indices represent
tensorial traces. These equations are referred to respectively as the Hamiltonian
constraint, momentum constraint and evolution equations.

\subsection{Conformal decomposition, maximal slicing and Dirac gauge}

Now we must choose a set of variables and a gauge, to get a partial
differential equations system that we
solve numerically. The first ingredient in the formalism presented in
\cite{Bon04} is the conformal decomposition of the three-metric~\cite{Lichn42}. We
define on each slice $ \Sigma_{t}$ an extra metric noted $f_{ij}$, that will
have a vanishing Riemann tensor (flat metric) and will be time
independent. The existence of such a metric in a neighborhood of
spatial infinity is ensured by our sub-manifold being
asymptotically flat. The associated flat connection is noted $
\mathcal{D}_{i}$. We introduce in $ \Sigma_{t} $ a 
conformal metric such that its determinant coincides with that of
$f_{ij}$, as:
\begin{equation}
\tilde{\gamma}_{ij} = \psi^{-4} \gamma_{ij};\quad \psi = \left(
\frac{\textrm{det}(\gamma)}{\textrm{det}(f)} \right)^{\frac{1}{12}}.
\label{e:defpsi}
\end{equation}
The tensor field $ h^{ij}$ we use to encode the conformal degrees of freedom
is the deviation of the conformal metric from the flat one:
\begin{equation}
 \tilde{\gamma}^{ij} = f^{ij} + h^{ij}.
\label{e:f+h}
\end{equation}

We also define in our equation sources the following conformal traceless extrinsic
curvature:
\begin{equation}
\hat{A}^{ij} = \psi^{10}(K^{ij} - \frac{1}{3}K\gamma^{ij});
\label{e:Aij}
\end{equation}
  
We choose for a gauge the generalized Dirac gauge for the conformal metric:
\begin{equation}
{\mathcal D}_{k} \tilde{\gamma}^{ki} = {\mathcal D}_{k} h^{ki} = 0,
\label{e:jauge}
\end{equation}
and we add to this prescription the maximal slicing condition, i.e the
vanishing of the trace in the extrinsic curvature: $ K = 0$. Therefore,
$\hat{A}^{ij}$ contains all the information about extrinsic geometry.
 
Under those conditions, we can rewrite the (3+1) Einstein equations
in what we shall call the FCF system:
\begin{eqnarray}
 \Delta \psi &=& \mathcal{S}_{\psi}(N, \psi, \beta^{i}, \hat{A}^{ij}, h^{ij}), \label{e:FCF1}\\
\Delta (N\psi) &=&  \mathcal{S}_{(N\psi)}(N, \psi, \beta^{i}, \hat{A}^{ij}, h^{ij}), \\
\Delta \beta^{i} + \frac{1}{3} \mathcal{D}^{i}\mathcal{D}_{j}\beta^{j}
&=&    \mathcal{S}_{\beta}^{i}(N, \psi, \beta^{i}, \hat{A}^{ij}, h^{ij}), \label{e:FCF2}\\
    \dder{h^{ij}}{t} - \frac{N^2}{\psi^4} \Delta h^{ij}
    &-& 2 \mathcal{L}_{\w{\beta}} \der{h^{ij}}{t} 
    + \mathcal{L}_{\w{\beta}}\mathcal{L}_{\w{\beta}} h^{ij} = \nonumber\\
     && \mathcal{S}_{h^{ij}}^{ij}(N, \psi, \beta^{i}, \hat{A}^{ij}, h^{ij}). 
\label{e:FCF}
\end{eqnarray}

 $\Delta$ is the usual scalar flat laplacian (which expression from a
spectral point of view is recalled in the Appendix~\ref{s:appendix}). The actual sources $\mathcal{S}_{\cdots}$, in general non-linear in the
variables and time-dependent, can be retrieved by the reader from \cite{Bon04}.
  
We must supplement this system with the kinematical relation between the
three-metric and extrinsic curvature of the slice, deduced from~(\ref{e:Kij})
and~(\ref{e:Aij}) (see equation~(92) of \cite{Bon04}).
%\begin{equation}
%    \hat{A}^{ij} = {\psi^{6}\over 2N} \left[ (L\beta)^{ij}
%        + \der{h^{ij}}{t} - \mathcal{L}_{\w{\beta}} h^{ij}
%        - {2\over 3} \cD_k\beta^k \, h^{ij} \right] . \label{e:Aij_calcul}
%\end{equation}
%Where $L$ is the conformal Killing operator linked to the flat three-metric.
This fully constrained scheme is strictly equivalent to the one presented in
\cite{Bon04}. A slightly different version has been presented recently in
\cite{Cord1}, focusing on non-uniqueness issues. Although the scheme
in \cite{Cord1} would probably pose
no additional difficulty in the present setting (except maybe some more
boundary conditions to prescribe to additional variables), there has been no
significant indication of problems involving non-uniqueness of solutions in
our study, that suggested modifications of the original formalism.

Here we choose as variables the quantities $ N \psi$, $\psi$,
$\beta^{i}$ and $h^{ij}$. We especially come up with three elliptic
equations, two scalar and one vectorial. Those are derived directly from
the Hamiltonian and momentum constraints of the (3+1) system, together
with the trace part of the dynamical equations. In an evolution scheme, these
will be the conditions enforced at each value of time
$t$. We do this for one particular slice. 

We are then left in general with a second-order tensorial hyperbolic
equation~(\ref{e:FCF}) dealing with the variable $h^{ij}$, that is obtained by the
geometrical relation between $\gamma_{ij}$ and $K_{ij}$, and the
dynamical part of Einstein equations. 

The goal here is to simulate as accurately as possible stationary spacetimes
containing one black hole, represented by an isolated horizon. In this
respect, we shall assume a coordinate system that is adapted to
stationarity. This will mean that a stationary timelike Killing vector field will be
identified with our time evolution vector field $(\frac{\partial}{\partial
  t})^{i}$. Using this prescription, all the time derivatives in our equations
vanish, so that our sources and operators simplify somewhat. In particular, the tensorial equation is written using an
  ellipticlike operator acting on $h^{ij}$:
\begin{equation}
\Delta h^{ij} -\frac {\psi^{4}}{N^{2}} \mathcal{L}_{\beta}
\mathcal{L}_{\beta}h^{ij} = \mathcal{S}_{2}^{ij}(h^{ij}, N, \psi, \beta,
A^{ij}).
\label{e:dh1}
\end{equation}
Our problem is then totally equivalent to an actual initial data problem,
where quantities have to be determined on a three-slice by elliptic equations,
before evolving them. The main difference with classical initial data schemes
like the conformal transverse traceless (CTT), the extended conformal thin sandwich (XCTS) scheme 
or the conformal flat curvature (CFC) system, is an additional elliptic equation for the conformal
geometry of the three-slice. Up to now, a vast majority of initial data
computations have been done using an {\it ad hoc\/} prescription for the
conformal geometry. The most common one is the conformally flat approach,
where $\tilde{\gamma_{ij}}$ is simply approximated to be the 3D flat metric. This
has been done in numerous computations, and this type of initial data is the
most frequently used for black hole evolution simulations. However, though
this conformally flat approximation turns out to be well-behaved in most
cases, we know that it is a strong limitation when trying to compute stationary
black hole spacetimes: it has been proven that the rotating Kerr-Newman
spacetime does not admit any conformally flat slice (see \cite{Val04b,Dainn}).

Other prescriptions for the conformal geometry include data suggested by the
post-Newtonian formalism \cite{PN}, or superposition of additional
gravitational wave content (see \cite{BSeid96}). Let us mention the work of
\cite{ShibLim} for neutron-star binary initial data, which also computes the
conformal geometry using a prescription in \cite{ShibUrFr}, that considers as
well the dynamical Einstein equations for the conformal variables. Finally,
the exact scheme we have explicated above has been applied by one of the authors in
the case of a single rotating neutron star in equilibrium \cite{LinNov06}. It
has led to the computation of strictly stationary initial data, that can be
directly extended into future and past time directions. This is exactly what
we are trying to do here in the black hole case.

\subsection{Resolution of conformal metric part}\label{ss:non_cf}

Apart from the boundary condition problem (that we discuss in Sec.~\ref{s:BCs}), our approach for the resolution of the tensorial
equation presents some peculiarities that we explain here.

The system of equations is composed of equation~(\ref{e:dh1}) and the gauge condition:
\begin{equation}
\label{e:probncf}
\mathcal{D}_{i} h^{ij} = 0,
\end{equation}
that we supplement with a condition on the determinant of
$\tilde{\gamma}^{ij}$, following from our definition of the conformal factor:
\begin{equation}
 \textrm{det}(\tilde{\gamma}^{ij}) = \textrm{det}(h^{ij} + f^{ij}) = 1.
\label{e:det1}
\end{equation}

We are left with a tensorial equation for a symmetric tensor with four
constraints: the system has two degrees of freedom. We now try to make them
explicit and solve for the related variables. Any second-rank symmetric tensor
$h^{ij}$ can be decomposed in the following way into a divergence free
part, and a symmetrized gradient part:
\begin{equation}
  h^{ij} = \mathcal{D}^{i}W^{j} +  \mathcal{D}^{j}W^{i} + h^{ij}_{T},
\label{e:hijtrans}
\end{equation}
with $\mathcal{D}_{i} h^{ij}_{T} = 0$. We shall use here variables associated
only with the divergence-free part $h^{ij}_{T}$, meaning that the gauge
component (gradient part) of the tensor considered has no influence on
them. We choose to encode the information in $h^{ij}_{T}$ in the two scalar
spectral potentials $A$ and $\tilde{B}$ presented in \cite{Cord2}, and whose
definitions are quickly recalled in the Appendix~\ref{s:appendix}. (A more
extensive study shall be performed in \cite{NovCV}).

What is remarkable about quantities $A$ and $\tilde{B}$ is that they can
actually be decomposed into scalar spherical harmonics, and that the tensorial
Poisson equation $\Delta h^{ij} = S^{ij}$ decouples into scalar elliptic
equations $A$ and $\tilde{B}$ (see the Appendix~\ref{s:appendix}). This is
not exactly the case for the nonlinear modified elliptic
operator~(\ref{e:probncf}); however, in our numerical scheme, we just slightly
modify the sources of the equation at each iteration so that we can write:
\begin{eqnarray}
\label{e:eqa1}
\Delta A -\frac {\psi^{4}}{N^{2}} \mathcal{L}_{\beta} \mathcal{L}_{\beta}A &=&
A_{\mathcal{S}}(h^{ij}, N, \psi, \beta, A^{ij}),\\
\label{e:eqb1} 
\tilde{\Delta} \tilde{B} -\frac {\psi^{4}}{N^{2}} \mathcal{L}_{\beta}
\mathcal{L}_{\beta}\tilde{B} &=& \tilde{B}_{\mathcal{S}}(h^{ij}, N, \psi,
\beta, A^{ij}),
\end{eqnarray}
the elliptic operator $\tilde{\Delta}$ being defined in the
Appendix~\ref{s:appendix}. We keep the Lie derivative notation for scalar
fields, to show that this is directly related to the operator in
Eq.~(\ref{e:dh1}); of course, in the scalar case, this operator simply reduces
to $ \mathcal{L}_{\beta}A = \beta^{i}\mathcal{D}_{i}A$. During the iteration
of the resolution algorithm, the sources of the equations are updated so that
they stay coherent with the original equation in $h^{ij}$. Equations
(\ref{e:eqa1}) and (\ref{e:eqb1}) are the two elliptic equations that we solve
at each iteration.

Specifically, at each step, we proceed as follows: once the scalars $A$ and
$\tilde{B}$ are determined by the resolution of ~(\ref{e:eqa1})
and~(\ref{e:eqb1}), the Dirac gauge and unit determinant conditions allow us
to totally reconstruct a divergence-free tensor, as the expected solution of
our tensorial equation. This is done by inverting two differential systems
((\ref{e:diracmag}) and (\ref{e:diracel}) of the Appendix~\ref{s:appendixB}),
that express the Dirac gauge conditions and definitions of the scalars $A$ and
$\tilde{B}$ in function of the tensor components. Those differential systems
involve scalars, which are components of $h^{ij}$ in a tensor spherical
harmonics basis (see the Appendix~\ref{s:appendix}, and Sec.~V of \cite{Cord2}).

The differential systems require three boundary conditions on the
excised surface, to be inverted (see \cite{NovCV}); we discuss them in Sec.~\ref{s:BCs},
in a detailed description of the scheme. In
addition, the trace of our tensor with respect to the flat metric is
kept fixed, so that the calculated
determinant at this step is one. This reduces to an algebraic
nonlinear condition for the tensor components. Finally, we update our sources
for the next step. We note here that the resolution for the variable $\tilde{C}$ introduced in the appendix is not necessary
in this scheme: for a divergence-free tensor, $\tilde{C}$ is unambiguously
determined by the knowledge of $\tilde{B}$ and the trace.

With this tensorial scheme, the gauge is necessarily enforced by construction,
so no gauge-violating mode can occur. This is in the same spirit as the global
fully constrained formalism (equations (\ref{e:FCF1}-\ref{e:FCF2})) for our
equation system, that forbids {\it a priori\/} all constraint-violating modes.
We also emphasize the fact that, in the general case and with an arbitrary
source for~(\ref{e:dh1}), we do not recover an actual solution of the
equation by reconstructing our tensor this way. This is only true if the
elliptic equation admits a solution that actually satisfies the Dirac gauge
and the determinant condition. We can see it as an integrability condition for
our equation, that is for example not generically true during an iteration.
However, since in our case we are looking for stationary axisymmetric data for
a single black hole, we know that our entire system does admit a solution: it
is the Kerr-Newman spacetime in Dirac gauge. As a consequence, if our scheme
converges, we know that the tensor field $h^{ij}$ we obtain shall satisfy the
dynamical Einstein equations, thus equation~(\ref{e:probncf}).
 
The missing ingredient for solving all our system of equations in an excised
spacetime is the setting of boundary conditions for our partial differential
equations, following part of the geometrical prescriptions of the isolated horizon
formalism, namely non-expanding horizon boundary conditions.

%%%%%%%%%%%%%%%%%%%%%%%%%%%%%%%%%%%%%%%%%%%%%%%%%%%%%%%%%%%%%%%%%%%%%%%%%%%%%%%%%
\section{Boundary conditions and resolution of the FCF system}\label{s:BCs}
%%%%%%%%%%%%%%%%%%%%%%%%%%%%%%%%%%%%%%%%%%%%%%%%%%%%%%%%%%%%%%%%%%%%%%%%%%%%%%%%%
\subsection{Boundary conditions for the constraint equations}\label{ss:BCcons}

Besides the prescription of asymptotic flatness at infinity
and the bulk stationarity prescription, all the physics of our system will be
contained in the boundary conditions we shall put on our excised
surface. This section follows largely the prescriptions of \cite{GJ06}.

We consider our excised two-surface to be a slice of an non-expanding horizon, i.e. a
MOTS with vanishing outgoing shear. Following Sec.~\ref{s:isol_hor},
this translates into several geometrical
prescriptions, namely the vanishing of the outgoing expansion and the shear two-tensor: $\theta_\ell =0$ and $\sigma_{ab}=0 $.
 
Being an instantaneous non-expanding horizon, the evolution of the
excision surface will be a null tube. Since we are adapting our coordinates to
stationarity, another important condition on the excised boundary
consists in prescribing the time evolution vector field of our
coordinates to be tangent to the null tube. 
Thus, we are ensured that our horizon location stays instantaneously fixed during
an evolution. Those prescriptions on the horizon will
suffice to give four boundary conditions for the constraint equations (one is scalar and the other vectorial), as we see
below.

We certainly have freedom to prescribe the coordinate location of our excision surface
in our coordinate system. For simplicity, we choose the surface to be a coordinate
sphere, fixed at a radius $r_{\mathcal{H}}$. We shall denote by $s^{i}$ the unit outer
spacelike normal to the surface, that will be tangent to the three-slice
$\Sigma_{t}$. The shift vector is then decomposed into two orthogonal parts
adapted to the geometry of the horizon: $ \beta^{i} = b s^{i} - V^{i}$.

The vanishing of the expansion can be expressed as a condition for the
conformal factor on the horizon:
\begin{equation}
4\tilde{s}^{i}\tilde{D}_{i}\ln(\psi) + \tilde{D}_{i}\tilde{s}^{i} +
\psi^{-2}K_{ij}\tilde{s}^{i}\tilde{s}^{j} = 0,
\label{e:thetaeq0}
\end{equation}
where we have used the conformal rescaling $\tilde{s}^{i} = \psi^{2} s^{i}$,
and the notation $\tilde{D}_{i}$ for the connection associated with the conformal
three-metric. Multiplying (\ref{e:thetaeq0}) by $\psi$, it 
can be seen as a non-linear Robin
condition for the quantity $\psi$.
The requirement for the time evolution vector field on the horizon to be tangent to
the null tube provides the equality $ b = N $. This is a natural way to fix the component of
the shift normal to the two-sphere.
% [ (DISCUTER POUR OMISSION DE CE PARAGRAPHE) However, some authors have expressed
%concerns about this boundary condition not being able to ensure uniqueness on the
%constraint system (see for example \cite{DainJK05}); a ``wrong sign'' on the
%quantity $ K_{ij}\tilde{s}^{i}\tilde{s}^{j}$ prevents the application of the
%maximum principle to the Hamiltonian constraint. However, numerical tests on
%an alternative prescription for $\tilde{b}$ \cite{JAL07} have not been really
%conclusive. We will then stick to a ``simple'' version here].
 We must also
fix  $V^{i}$, the part of the shift tangent to the two-surface. For
this we make use of the vanishing of the symmetric shear tensor $\sigma_{ab}$. It can be shown
(\cite{CookPf04} and \cite{GJ06}) that the vanishing of the shear is equivalent
to the following equation for $V^{i}$:
\begin{equation}
q_{bc} \prescript{2}{}{D}_{a}V^{c} +
 q_{ac} \prescript{2}{}{D}_{b}V^{c} -
q_{ab} \prescript{2}{}{D}_{c}V^{c} = 0.
\label{e:sheareq0} 
\end{equation}

Here $\prescript{2}{}{D}$ is the connection associated with $q_{ab}$ on the
surface. This means that $V^{i}$ is a conformal Killing symmetry for the
two-sphere (in particular, quantities in (\ref{e:sheareq0}) can be substituted
by {\it tilded} conformal ones). Defining coordinates $(\theta, \varphi)$ on
our two-sphere, we prescribe $V^i$ as:

\begin{equation}
V^{i} = \Omega \left(\frac{\partial}{\partial \varphi}\right)^{i},
\label{e:ckv}
\end{equation}
and we shall verify {\it a posteriori} that this is a (conformal) axial symmetry. The constant $\Omega$ will be called the rotation rate of the horizon, $\varphi$
being the azimuthal coordinate. In the case of the Kerr spacetime,
there is an analytical relation between the areal radius of the
apparent horizon, the (reduced) angular momentum parameter
$\frac{a}{M}$, and $\Omega$. From
a more general point of view, different values for $\Omega$ will
likely affect directly the angular momentum.
In the general case, we define a parameter $a$ for the angular momentum
associated with the entire
spacetime, from the dimensionless relation:
\begin{equation}
\frac{a}{M_{ADM}} =
\frac{J_{K}}{M_{ADM}^{2}}
\label{e:angular_param}
\end{equation}
with $M_{ADM}$ the ADM mass of the 3-slice,
and $J_{K}$ the Komar angular momentum of the 3-slice at infinity; the latter is tentatively defined with
the (presumably) Killing vector $\left(\frac{\partial}{\partial
    \varphi}\right)^{i}$ (see Equations (7.14) and (7.104) in \cite{Gourg07} for explicit expressions for $M_{ADM}$ and $J_{K}$). Note that we do not impose any Killing symmetry, except on the
horizon: we know however, by the black hole rigidity theorem \cite{HawEll73},
that an accurate resolution of Einstein equations would impose this vector to
be so. We discuss the dependence between all those quantities in Sec.~\ref{s:resus}.

Once we have set boundary conditions for the conformal factor and the three
components of the shift vector, we must still fix the lapse
function on the horizon. Different prescriptions have been considered 
in the literature (e.g. \cite{Cook02,CookPf04,CauCooGri06,GJM04,JAL07,GJ06}).
In the spirit of the effective approach in 
\cite{CookPf04,CauCooGri06}, we arbitrarily impose the value of the
lapse to be a constant $N_{\mathcal{H}}$ on the excised sphere.

As mentioned before, previous boundary conditions define with no ambiguity our
excised surface to be a slice of a non-expanding horizon. Moreover, our choice
for the lapse, the horizon location and the conformal Killing symmetry on the
horizon fixes coordinates on the two-surface. Only the conformal two-geometry
of the excised sphere remains to be fixed. This is done in relation with the
resolution scheme for the $h^{ij}$ equation.

\subsection{Boundary conditions for the $h^{ij}$ equation}\label{ss:BChij}

We recall that, with the approach developed in Sec.~\ref{s:FCF}, the
resolution of our tensorial problem in Dirac gauge reduces to two elliptic-like
scalar equations, to be solved on a three-slice excised by a two-sphere: we should
normally provide two additional boundary conditions for those equations.

A result by \cite{Cord3} shows that in the full evolution case for this tensorial
equation~(equation~(\ref{e:FCF})) and in a Dirac-like gauge, the characteristics of the
equation are not entering the resolution domain when the spacetime is excised by a null or spacelike
marginally trapped tube. This means that in the evolution case, once the
initial data are set, there is no boundary condition whatsoever to prescribe to
the hyperbolic equation.

The problem is of course different here, where we are left with
an elliptic equation instead of a hyperbolic one. However, a simple analysis
will hint that in our particular single horizon case, there will not be any inner
boundary condition to be prescribed on our data.

Let us examine the case of the elliptic equation in $A$, that we recall here:
\begin{equation}
\Delta A -\frac {\psi^{4}}{N^{2}} \mathcal{L}_{\beta} \mathcal{L}_{\beta}A =
A_{\mathcal{S}}(h^{ij}, N, \psi, \beta, \hat{A}^{ij}).
\label{e:opa}
\end{equation}

We will try and exhibit a simplified linear operator acting on the variable
$A$, that will contain the most relevant terms. The double Lie derivative
operator acting on $A$ can be separated in:
\begin{equation}
\frac {\psi^{4}}{N^{2}} \mathcal{L}_{\beta} \mathcal{L}_{\beta}A =
\frac{\psi^{4}}{N^{2}}(\beta^{r})^{2}\partial_{r}^{2}A + \frac
     {\psi^{4}}{N^{2}}( \mathcal{L}_{\beta} \mathcal{L}_{\beta}A)^{*};
\label{e:seplie}
\end{equation}
the second term contains all the remaining components of the double Lie derivative.

At this point, and with a fixed system of spherical coordinates, we are allowed
to make a decomposition into spherical harmonics for all the scalar variables.
We write in this respect: 
\begin{equation}
 A = \underset{(l, m)}{\sum} A_{l m} Y_{lm}(\theta, \varphi),
\label{e:alm}  
\end{equation}
where $Y_{l m}$ are the spherical harmonics of order $(l, m)$, defined as
eigenfunctions of the angular Laplace operator: $\Delta_{\theta
  \varphi}Y_{l m} = -l(l +1) Y_{l m} $. 

We now point out the fact that, due to our coordinate choice, we have
$(\beta^{r})_{(l=0)} = (\frac{N}{\psi^{2}})_{(l=0)}$ on the horizon. We can use a second-order Taylor
expansion to write the $l =0$ part of the factor in front of the first term of~(\ref{e:seplie}),
close to our surface coordinate radius $r_{\mathcal{H}}$:
\begin{eqnarray}
\label{e:opmoda}
 \left[\frac{\psi^{4}}{N^{2}}(\beta^{r})^{2}\right]_{(l=0)}\partial_{r}^{2}A =
 [1 + \alpha(r -r_{\mathcal{H}}) + \nonumber \\
 \delta(r-r_{\mathcal{H}})^{2} + \mathcal{O}(r-r_{\mathcal{H}})^{3}]\partial_{r}^{2}A,
\end{eqnarray} 
where $\alpha$ and $\delta$ are two real numbers that can be directly computed
during one iteration, from the values of $N$, $\psi$ and $\beta^{r}$ at the excised
surface. Our global equation can be rewritten for each spherical harmonic $l$
as:
\begin{eqnarray}
\label{e:opmodalm}
\left[ -\alpha(r -r_{\mathcal{H}}) - \delta(r-r_{\mathcal{H}})^{2} \right]
\frac{\partial^{2}}{\partial r^{2}}A_{l m} + \frac{2}{r}
\frac{\partial}{\partial r}A_{l m} \nonumber \\
 -\frac{l(l+1)}{r^{2}}A_{l m}
= A_{\mathcal{S}}+  \frac {\psi^{4}}{N^{2}}( \mathcal{L}_{\beta}
\mathcal{L}_{\beta}A)^{**}_{l m},
\end{eqnarray}
where we only keep on the left-hand side the terms given in~(\ref{e:opmoda}),
and put the rest (denoted with  ${}^{**}$) with the source. The latter contains the remaining
components of the double Lie derivative, and involves either  terms that are
not second-order in the radial derivative, or that are multiplied by the higher
harmonics of $\frac{\psi^{4}}{N^{2}} (\beta^{r})^{2}$ (supposedly smaller than
the main term, explicitly developed in (\ref{e:opmodalm})). Thus, we have isolated a linear
operator $\mathcal{Q}_{\alpha \delta}$, depending on two real numbers $\alpha$
and $\delta$:
\begin{eqnarray}
\label{e:opQ}
\mathcal{Q}_{\alpha\delta} &=&\left[ -\alpha(r -r_{\mathcal{H}}) - \delta(r-r_{\mathcal{H}})^{2}\right]
\frac{\partial^{2}}{\partial r^{2}} \nonumber \\
 &+& \frac{2}{r}
\frac{\partial}{\partial r} -\frac{l (l +1)}{r^{2}}I,
\end{eqnarray}
other contributions being taken as source terms. This operator is different
from the ordinary Laplace operator by the factor in front of the second order
differential term, which vanishes on the excision boundary. It can be shown
that the space of analytic solutions on $\mathbb{R}^{3}$ minus the excised
horizon belonging to the kernel of $\mathcal{Q}_{\alpha\delta}$ is generally of dimension one. This is in contrast with the case
of the Laplace equation, where it is of dimension two. In practice, this will
mean that for a numerical resolution of an equation $\mathcal{Q}_{\alpha
  \delta} f = S_{f}$, there is only one boundary condition to fix for the
unknown, mainly the behavior at infinity. No additional information is needed
at the excised boundary for the {\em effective}
operators (\ref{e:opQ}). Operators of this kind are known in the mathematical
literature as elliptic operators with weak singularities \cite{HNW}.

 The very same scheme can be applied to the equation~(\ref{e:eqb1}) for
 $\tilde{B}$, the only difference being that the original Laplace
 operator is replaced by a slightly modified one (see the Appendix~\ref{s:appendix}).

 As explained in Sec.\ref{ss:non_cf}, after solving for the two main equations
 (\ref{e:opmodalm}) and its equivalent in $\tilde{B}$, the inversion of the
 gauge differential systems (\ref{e:diracmag}-\ref{e:diracel}) (explicited in the
 appendix~\ref{s:appendixB}) for the reconstruction of $h^{ij}$ requires three
 extra boundary conditions, in addition to the vanishing of all quantities at
 infinity. We obtain them as compatibility conditions based on the original
 elliptic tensorial equation~(\ref{e:dh1}): we express three decoupled
 elliptic scalar equations for three components of $h^{ij}$ in the
 spin-weighted tensor spherical harmonics basis,denoted $h^{rr}$, $\eta$ and $\mu$, which are
 directly related to the usual tensorial components of $h^{ij}$ and defined in
 Appendix~\ref{s:appendix}. From the tensor equation~(\ref{e:dh1}), we deduce:
\begin{eqnarray}
\Delta \mu +\frac{2}{r}\frac{\partial
    \mu}{\partial r} + \frac{2\mu}{r^2} - \frac{\psi^{4}}{N^{2}} \left(\mathcal{L}_{\beta}
\mathcal{L}_{\beta}h^{ij}\right)^{\mu} = \left(S_{2}^{ij}\right)^{\mu}
 \text{       } \label{e:ellmu}\\
  \Delta \eta + \frac{2}{r}\frac{\partial
    \eta}{\partial r} + \frac{2\eta}{r^2} + \frac{2 h^{rr}}{r^{2}} - \frac{\psi^{4}}{N^{2}} \left(\mathcal{L}_{\beta}
\mathcal{L}_{\beta}h^{ij}\right)^{\eta} =
\left(S_{2}^{ij}\right)^{\eta}\text{       } \label{e:elleta}\\
\Delta h^{rr} - \frac{6h^{rr}}{r^2}
  -\frac{4}{r^2}\lapang \eta  + \frac{2h}{r^2}
 - \frac{\psi^{4}}{N^{2}} \left(\mathcal{L}_{\beta}
\mathcal{L}_{\beta}h^{ij}\right)^{rr} \nonumber \\
=  S_{2}^{rr} , \label{e:ellrr}\text{       }
\end{eqnarray}
where $(\mu, \eta, rr)$ superscripts indicate the corresponding components of
$h^{ij}$ in the tensor spherical harmonics basis (see
Appendix~\ref{s:appendix}). As for the equation involving $A$, we can rewrite
the above equations by extracting the weakly singular operator
$\mathcal{Q}_{\alpha\delta}$ acting on the principal variable, the other
contributions being put on the right-hand side of the equations. For example,
the equation in $\mu$ can be rewritten the following way:
\begin{equation}
 \mathcal{Q}_{\alpha\delta}(\mu) +\frac{2}{r}\frac{\partial
    \mu}{\partial r} + \frac{2\mu}{r^2} =
 \left(S_{2}^{ij}\right)^{\mu} + \frac{\psi^{4}}{N^{2}} \left(\mathcal{L}_{\beta}
\mathcal{L}_{\beta}h^{ij}\right)^{\mu (**)},
\end{equation}
 with the Lie derivative term containing all the left-hand-side
 contributions of equation~(\ref{e:ellmu}) not taken into account. We do not need to invert this equation: however, as the leading order
 term in $\mathcal{Q}_{\alpha\delta}$ vanishes at the horizon
 $(r_{\mathcal{H}}=1)$, we can accordingly write a Robin-like boundary condition for
 the $\mu$ quantity:
\begin{equation}
\label{e:boundmu}
\frac{4}{r}\frac{\partial
    \mu}{\partial r} + \frac{(\lapang + 2)}{r^2}\mu  =  \left(S_{2}^{ij}\right)^{\mu} + \frac{\psi^{4}}{N^{2}} \left(\mathcal{L}_{\beta}
\mathcal{L}_{\beta}h^{ij}\right)^{\mu (**)}.
\end{equation}

This will be used as a boundary condition for the gauge system (\ref{e:diracmag}),
the source terms being computed with quantities from the previous
iteration. Using the same method, we can write very similar expressions (that
we do not explicitly give here) for the fields $h^{rr}$ and $\eta$, to be used
as Robin boundary conditions applied to the gauge differential system
(\ref{e:diracel}). The three boundary conditions are sufficient to invert the
two gauge systems (\ref{e:diracmag}-\ref{e:diracel}) \cite{NovCV}, and reconstruct
the whole $h^{ij}$ tensor from the tensor spherical harmonics components (see
the Appendix~\ref{s:appendixB} and \cite{NovCV} for details).

To summarize, the method employed for the resolution of the whole $h^{ij}$ system is
iterative and can be decomposed for each step in the following way (more
technical details are provided in the Appendices):
\begin{enumerate}
\item After calculating the source $\mathcal{S}_{2}^{ij}$ from
  equation~(\ref{e:dh1}), we deduce the right hand side of the
  equation~(\ref{e:opmodalm}) for $A$, using values from the previous
  iteration. The same is done for the quantity $\tilde{B}$ and its
  corresponding source terms.
\item We invert equations (\ref{e:opmodalm}) and its equivalent for
  $\tilde{B}$, only by imposing that the fields are vanishing at
  infinity.
\item We compute the value of the trace from equation~(\ref{e:det1}) (a more
  explicit expression can be found in equation~(169) of \cite{Bon04}). This
  allows us to write the two differential systems (Dirac gauge systems) mentioned in
  Sec.~\ref{ss:non_cf} and expressed in Appendix~\ref{s:appendixB}, involving
  the spherical harmonics components of $h^{ij}$ (scalar quantities).
\item We invert these two gauge differential systems using three boundary conditions
  similar to Eq.~(\ref{e:boundmu}), for the three scalar spherical harmonics
  components $h^{rr}$, $\eta$ and $\mu$. As those are compatibility conditions
  expressing information already contained in Eq.~(\ref{e:dh1}), we provide in
  this way no additional physical information. This
  gives us the spherical harmonics components of $h^{ij}$.
\item We reconstruct the whole tensor $h^{ij}$ from the spherical
  harmonics components. 
\end{enumerate}

%detailed in \ref{s:appendixB}
%This is the method we use to solve the whole $h^{ij}$ system, without any
%additional prescription on the horizon. At a particular step, we invert
%operators in~(\ref{e:eqa1}) and~(\ref{e:eqb1}) to get $A$ and
%$\tilde{B}$. From these variables, as mentioned in Sec.~\ref{s:FCF}, we
%compute the trace by the determinant condition, and we retrieve the other
%tensor components by inverting a differential system that expresses the Dirac
%gauge conditions and the definitions of $A$ and $\tilde{B}$, as mentioned in
%Sec.~\ref{ss:non_cf}. This differential system needs three additional
%conditions, which will be prescribed as compatibility conditions : 
%we can express,
%for the tensor harmonics components of $h^{ij}$, three decoupled elliptic
%equations, namely for $h^{rr}$ and the potentials $\mu$ and $\eta$ defined in the appendix,
% involving again the operator $\mathcal{Q}_{\alpha \delta}$. Since the coefficient of 
%leading order for this operator is vanishing at the boundary, the elliptic equations 
%amount to Robin boundary conditions for the three components at the horizon.Those are
%sufficient to invert the differential system. In this way, we add no
%additional information to reconstruct $h^{ij}$: only information already
%contained in the tensorial equation is needed. {\bf REVOIR}

We have not proven here that no boundary condition has to be put generically for
the resolution of the two scalar equations involving $A$ and $\tilde{B}$ in
the tensorial system. However, if we implement numerically the resolution by
the inversion of the operator $\mathcal{Q}_{\alpha \delta}$ at each
iteration, we will not have to impose any boundary condition, but only
informations coming from the Einstein equations. Moreover, a
convergence of the entire $h^{ij}$ system would support the coherence of the
reasoning, and hint that there is, in our case, a deeper physical motive preventing the
prescription of additional information on the horizon. The results
in Sec.~\ref{s:resus} show this is the case.

%%%%%%%%%%%%%%%%%%%%%%%%%%%%%%%%%%%%%%%%%%%%%%%%%%%%%%%%%%%%%%%%%%%%%%%%%%%%%%%%% 
\section{Numerical results and tests}\label{s:resus}
%%%%%%%%%%%%%%%%%%%%%%%%%%%%%%%%%%%%%%%%%%%%%%%%%%%%%%%%%%%%%%%%%%%%%%%%%%%%%%%%%
\subsection{Setting of the algorithm}

All the numerical and mathematical tools we use here are available in the open
numerical relativity library LORENE \cite{LORENE}. Our simulation is made on a
3D spherical grid, using spherical harmonics decomposition for the angular
part and multidomain tau spectral methods (see \cite{GrandNov} for a review).
The mapping consists in four shells and an outer compactified domain, so that
infinity is part of our grid and we have no outer boundary condition to put at
a finite radius. Our grid size is typically $ N_{r}\times N_{\theta}\times
N_{\varphi} = 33\times 17\times 1$. We also have checked our code by setting
$N_{\varphi} = 4$, to verify that no deviation from axisymmetry occurred. Our
innermost shell has a boundary at the radius $r_{\mathcal{H}}$, which will be the
imposed location of a MOTS, and will be used as the unit of length in
all the results presented here. We impose the values of all the fields to be
equivalent at infinity to those of a flat three-space. Finally, trying to get
stationary data, we prescribe our coordinates to be adapted to this
stationarity, so that all the time derivatives in the Einstein (3+1) system are
set to zero. However, even if we expect to get axisymmetric data (the only
vacuum stationary solution for a black hole being the Kerr solution), we are
always able to solve our equations in three dimensions.

We proceed with our scheme in the following way: during one iteration, all the
variables are updated immediately after they have been calculated, so
that the sources for the next equations are modified. The tensorial
equation for $h^{ij}$ is the last solved in a particular iteration, and we obtain
at each step a
local convergence for the whole tensorial system (including the
determinant condition), before we proceed to the update of all
quantities, and to the next iteration.
   
We impose on the sphere of radius $r_{\mathcal{H}}$ the conditions of zero
expansion~(\ref{e:thetaeq0}) and shear~(\ref{e:sheareq0}), via respectively a
Robin condition on the quantity $\psi$ and a Dirichlet condition on the
partial shift $V^{i}$. We also impose the horizon-tracking coordinate
condition on the radial shift component~$ b $. Having set the shape and the
location of the surface in our coordinates, we are only left with two free
parameters, which are the boundary value of the lapse function and the
rotation rate $ \Omega$. As we said, the lapse function, which is
merely a slicing gauge choice, is fixed to a constant value $0 <
N_{\mathcal{H}} < 1$ on the horizon. We generate two sets of data on our
three-slice, spanning the rotation rate from zero (Schwarzschild solution) to
a value of about $0.22$, where our code no longer converges. One set will give
the solution for the whole differential system (the non-conformally flat (NCF)
data, supposed to converge to the rotating Kerr solution), while the other
will compute conformally flat (CF) data, by putting $h^{ij}=0$. From a
spacetime point of view, the CF data can also be seen as a computation of
black hole spacetime using the so-called Isenberg-Wilson-Mathews approximation
to general relativity \cite{IWM}, \cite{IWM2}.

\subsection{Numerical features of the code}
  
\begin{figure}
\centering
\includegraphics*[width= 8cm]{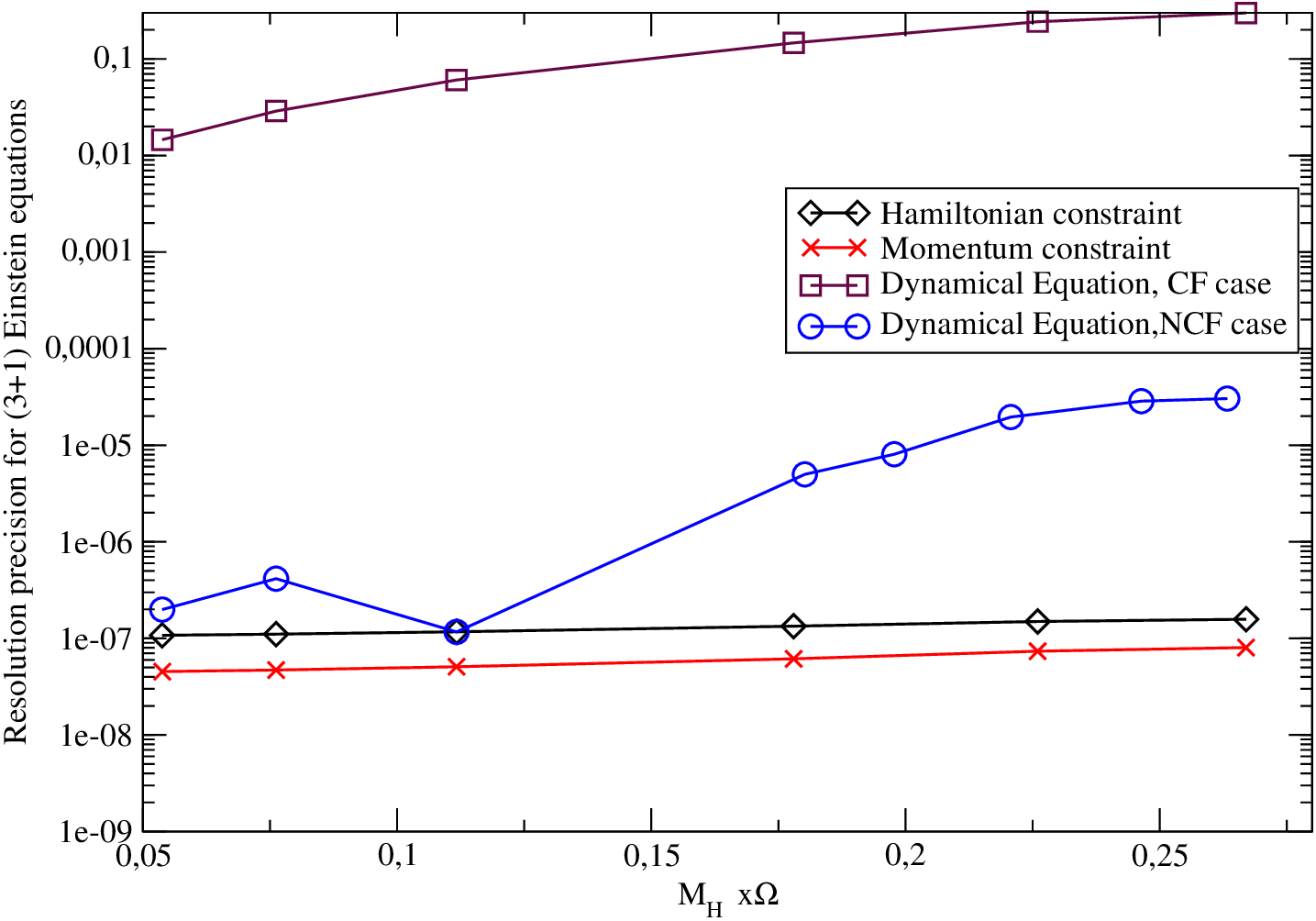}
\caption{Accuracy for Einstein equations resolution (in the original
  (3+1) version of \cite{York79}) as a function of
  dimensionless parameter $M_{\mathcal{H}}\Omega$ (see
  section~\ref{ss:pgtests} for a definition of $M_{\mathcal{H}}$). Data are absolute maximum
  error values for both constraint equations, and dynamical equations in both
  cases. Data are taken with $ N_{r}= 33$, $N_{\theta}=17$, $N_{\varphi}=1$.
  Lapse on the horizon is $N_{\mathcal{H}} =0.55$.}
\label{f:precision}
\end{figure}

Figure~\ref{f:precision} presents, on the one hand, the absolute accuracy
obtained for the Einstein constraints (in the form expressed in
\cite{York79}) in the NCF case. Regarding fulfillment of the Einstein
dynamical equation, Figure~\ref{f:precision} also shows the accuracy of the
NCF fully stationary solution, as well as its violation in the conformally
flat case. We see the expected improvement for precision of resolution of
dynamical equations in the full NCF case. Let us note that a verification of
the gauge conditions is not even necessary, as it is fulfilled by construction
(we only solve for variables satisfying the gauge). This is one of
the strengths of our algorithm.

\begin{figure}
\centering
\includegraphics*[width = 8cm]{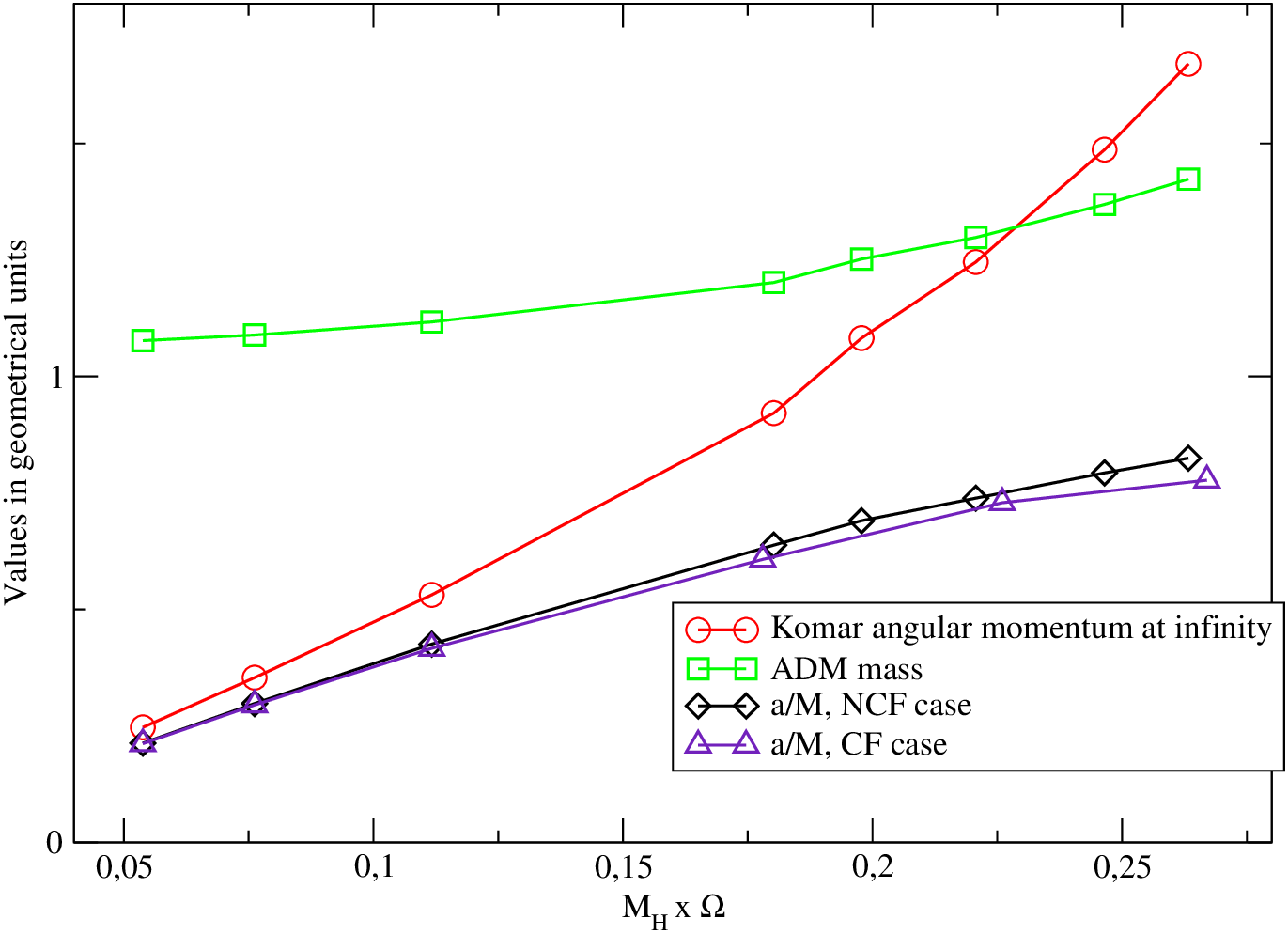}
\caption{Dependence on the parameter $M_{\mathcal{H}}\Omega$ of the ADM mass, the
  Komar angular momentum at infinity $J_{K}$ and the angular momentum
  parameter $\frac {a}{M}$ (both defined in section \ref{ss:BCcons}) for
  both cases. The value of the lapse on the horizon is here fixed at
  $N_{\mathcal{H}} = 0.55$.}
\label{f:omega_asM}
\end{figure}

A non-trivial issue of our computation is the link between the two
physical characteristics of the system (the mass and angular momentum of the
data) and the two input quantities supposed to fix them, namely the boundary
value for the lapse and the rotation rate $ \Omega $ on the horizon. We
choose here in our sequence to fix the value of the horizon coordinate radius,
removing it from the list of variables. Results are shown in figure~\ref{f:omega_asM}. The value of the lapse being also
fixed, we observe that an increase in $ \Omega $ not only affects the angular momentum, but
also the ADM mass of the spacetime. Moreover, fixing the rotation rate does not amount  to the
prescription of  the angular
momentum to an {\it a priori\/} given value. 
A decrease in the value of $N_{\mathcal{H}}$ on the horizon results
also in an increase in $\frac{a}{M}$ (defined in section \ref{ss:BCcons}). This stems from the fact that
our choice for the slicing directly influences in this approach the physical parameters
(e.g the areal radius)
of the solution obtained. We note also that for a fixed value of $N_{\mathcal{H}}$, the
correspondence between $\Omega$ and $\frac {a}{M}$ is slightly different in
the conformally flat case and in the NCF case.
With our algorithm, a larger value of the lapse gives a slightly better
convergence of the code for high rotation rates of the black hole (until
$N_{\mathcal{H}} =0.8$ approximatively). For each lapse the code stops converging at a
certain value of the rotation rate. We do not yet know whether this is a
problem of our algorithm to be improved, or if this has deeper physical
reasons: constant values for the lapse and the rotation rate might not
be ``good'' variables for the Kerr black hole in Dirac gauge, once we reach
high rotation rates. The only conclusion we can draw from this is that there
is a non trivial correspondence between our ``effective parameters'' $N_{\mathcal{H}}$ and
$\Omega$, and the physical ones, namely the ADM mass and Komar angular
momentum. This correspondence is likely to be one to one for values of $\frac
{a}{M}$ below a certain threshold of about $0.85$. Reaching higher
values for $\frac {a}{M}$ is left to future numerical investigations.

Let us mention again the remark made by \cite{CookPf04} about the boundary
condition for the lapse in the XCTS scheme. Although it is necessary to fix
the slicing of the spacetime by an arbitrary boundary condition on the lapse,
we have the freedom to decide what kind of condition to impose. The authors in
\cite{CookPf04} suggest that an arbitrary condition of Neumann or Robin type
would be preferable, because it is more flexible in view of a numerical algorithm.
In particular, not fixing a value for the lapse on the horizon, but rather
giving a first order prescription, allows the data to ``adapt'' to potentially
high tidal distortions. However, having also tried to impose Neumann
conditions for the lapse in our configurations, we do not see any clear
improvement in the robustness of the algorithm. This is why we still keep a
Dirichlet boundary condition as the simplest prescription.

\subsection{Physical and geometrical tests for stationarity}\label{ss:pgtests}
 
\begin{figure}
\centering
\includegraphics*[width = 8cm]{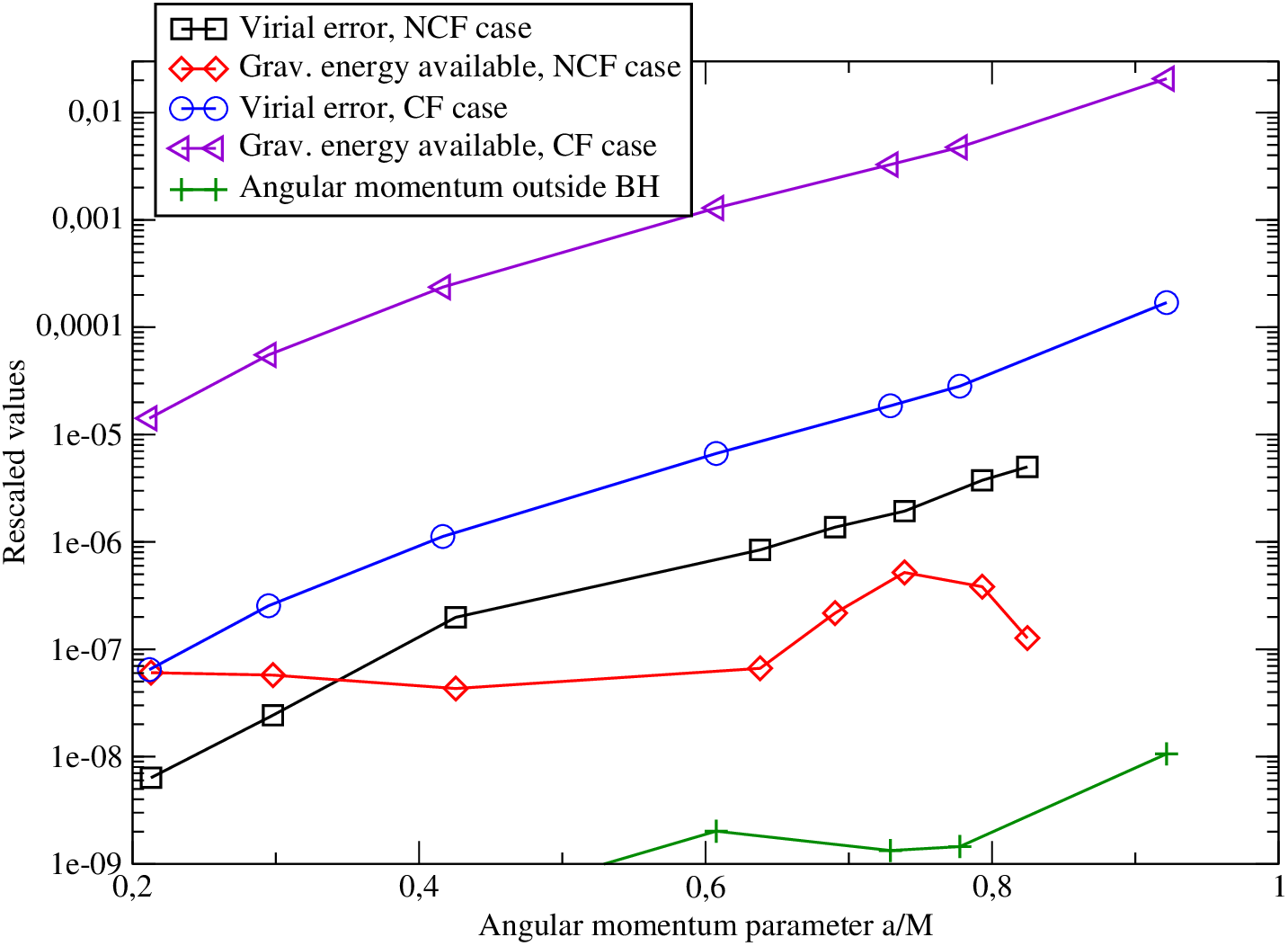}
\caption{Different diagnostics for stationarity in both cases, comparing
  physical quantities at the horizon and at infinity. The virial error
  computes the difference between ADM and Komar masses at infinity,
  rescaled with the ADM mass, and using
  asymptotic behaviors of the lapse and conformal factor. The radiation energy content outside the black hole (resp. outer angular momentum content) is the
  absolute difference between the horizon mass $M_{\mathcal{H}}$ (resp. angular momentum on the
  horizon $J_{\mathcal{H}}$) and the ADM mass $M_{ADM}$ (resp. Komar angular momentum at
  infinity $J_{K}$), rescaled with the ADM mass (resp. Komar angular momentum
at infinity).}
\label{f:error_JM}
\end{figure}

One of the tests of stationarity to be made can be the comparison between the ADM
mass and the Komar mass at infinity, defined with the (presumably) Killing vector $\left(\frac{\partial}{\partial
    t}\right)^{i}$ (equation~(7.91) of \cite{Gourg07}). The results of this test are displayed in figure~\ref{f:error_JM}. The
comparison between the ADM mass and the Komar mass is actually directly linked
to the Virial theorem of general relativity put forth by \cite{Virial94}. The
concordance between those masses is equivalent to the vanishing of the Virial
integral, and has been also used as a stationarity marker by \cite{GGB02}.

We have also computed in both NCF and CF cases an estimate of the amount of gravitational radiation
contained outside the black hole in the 3-slice. Following the prescriptions of
\cite{AshBF}, we calculate the
difference between the ADM mass and what could be called the isolated horizon
mass, defined in geometrical units by:
\begin{equation}
\label{e:BHmass}
 M_{\mathcal{H}}=\frac{\sqrt{R_{\mathcal{H}}^{4} + 4J_{\mathcal{H}}^{2}}}{2R_{\mathcal{H}}}.
\end{equation}
where $R_{\mathcal{H}}$ is the areal radius of $\mathcal{H}$. $ M_{\mathcal{H}}$ is nothing
but the formula for the Christodoulou mass \cite{Christ} calculated from the
Komar angular momentum $ J_{\mathcal{H}} $ on the horizon (defined
with the same supposed Killing symmetry as $J_{K}$). 
If we have an
isolated horizon extending to future infinity,  the difference
between  $ M_{ADM}$ and  $ M_{\mathcal{H}}$ gives exactly the
radiation energy emitted at future null infinity for the data \cite{AshBF}.
In non-stationary cases (for example binary systems), this is an appropriate estimate of the 
radiation content at an initial given time.

Results for comparison between the two cases studied here are shown in
figure~\ref{f:error_JM}. Although the gravitational energy available for NCF
spacetimes is contained under $10^{-7}$ whatever the rotation rate might be,
in the CF case, the increase in energy with $ \frac{a}{M}$ is patent. This
measure of energy available with respect to $\frac{a}{M}$ gives us a way of
approximating {\it{a priori\/}} the amount of what is usually called ``junk''
gravitational radiation, that could be emitted on a spacetime evolution with
conformally flat initial data. 

\begin{figure}
\centering
\includegraphics*[width = 8cm]{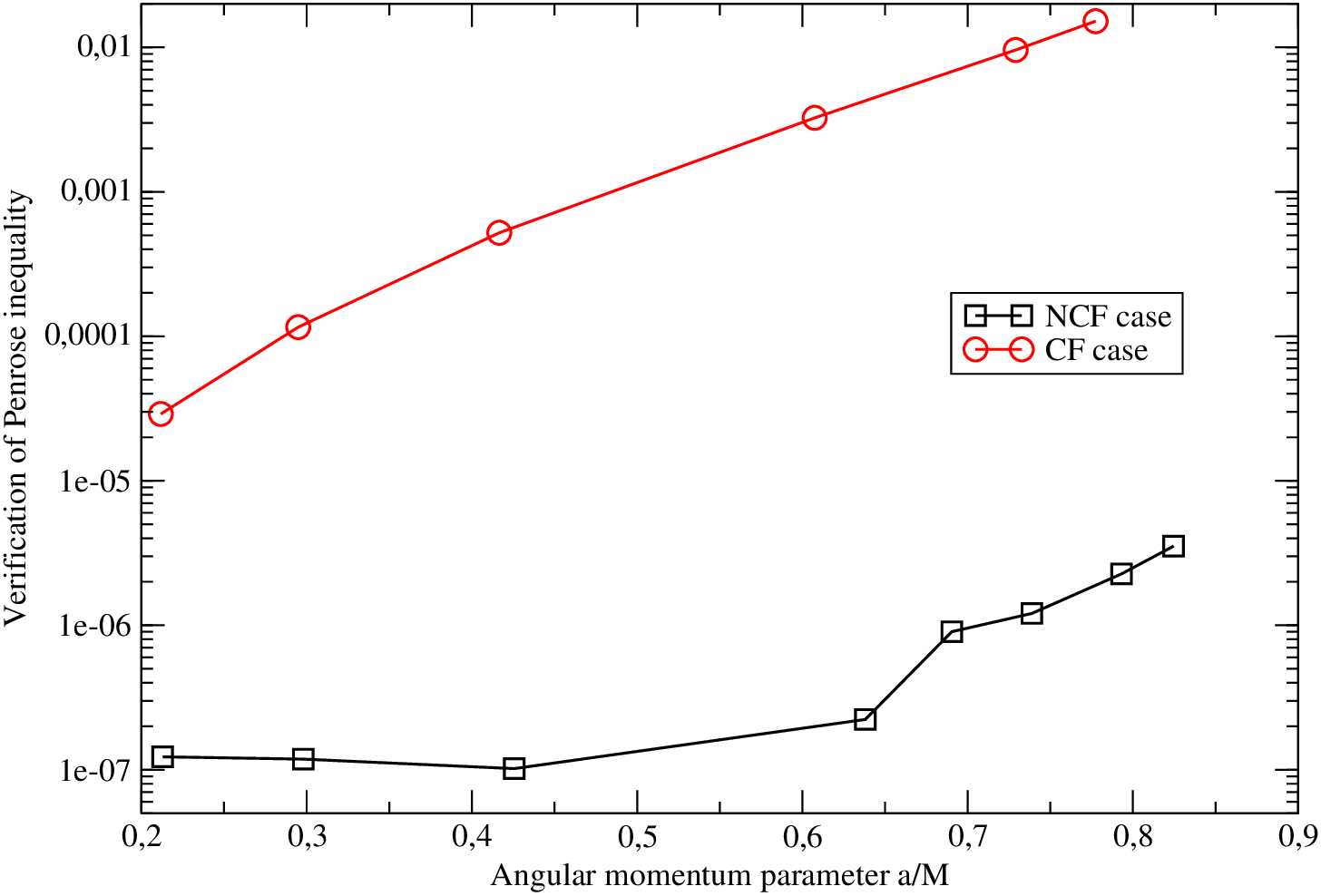}
\caption{Value of $1 - \epsilon_{A}$ for both data sets.}
\label{f:Epsilon}
\end{figure}

In the same spirit, we have also computed the accuracy in the verification of a Penrose-like
inequality for axisymmetric data, that can be written as:
\begin{equation}
\label{e:dain}
  \epsilon_{A} = \frac{\mathcal{A}}{8 \pi (M_{ADM}^{2}+ \sqrt{M_{ADM}^4 -
      J_{K}^{2}})} \leq 1,
\end{equation}
where $\mathcal{A}$ is the minimal area of a surface containing the horizon, $J_{K}$
is the Komar angular momentum at infinity  
and $M_{ADM}$ is the ADM mass at infinity. Being a little more stringent that the actual
Penrose inequality, it has been first proposed by \cite{HawEll73} for
axisymmetric spacetimes. This inequality is supposed to be verified for all
axisymmetric data containing an apparent horizon, and to be an equality only for
actual Kerr data (this is
referred in \cite{JAV07} as {\em Dain's rigidity conjecture} \cite{DaiLouTak02}). The results are presented in
figure~\ref{f:Epsilon}. We observe that, if the equality is very well verified
in the actual Kerr case, this is definitely not true for CF data, even for
reasonable values of $\frac{a}{M}$. In \cite{JAV07} (cf. \cite{JarVal07} for a general context), it has been proposed that this
quantity $ \epsilon_{A}$ ({\em Dain's number}) should be understood as a strong diagnosis tool for
distinguishing between Kerr horizons and other isolated or dynamical horizons.
This numerical observation shows strong support in favor of this claim,
pulling apart actual Kerr data and reasonable approximations of these data.
Let us also point out the virtual costlessness of this tool, as we only have
to rely on a single real value.

We also note that, when computing the
rescaled difference of Komar angular momentum between the horizon and
infinity $\frac{J_{K} - J_{\mathcal{H}}}{J_{K}}$, we come
up in all cases with a difference at the level of numerical precision for
resolution (see figure \ref{f:error_JM}). This is of course coherent with the fact that gravitational waves cannot carry any angular
momentum in axisymmetric spacetimes. This result ensures us the equivalence in practice between
the estimation of radiation exterior to the horizon and the verification of Penrose
inequality via Dain's number.

\subsection{Multipolar analysis}\label{ss:multipoles}

To be much more complete about the geometry of the constructed horizons, one
could rely on the source multipole decomposition of the two-surface lying on our
three-slice. This feature has first been presented by \cite{AshENPVDB}, based on
an analogy with electromagnetism, and first studied in \cite{SchKriBey06} in the
case of dynamical horizons. We here implement the computation of multipole
moments in the isolated horizon case, which is the strict situation where they
have been defined in \cite{AshENPVDB}.

A prerequisite is the existence of a preferred divergence-free vector field  $\varphi^a$ on
the sphere, from which the angular momentum of the horizon is defined (the
divergence-free condition on $\varphi^a$ ensures that all definitions will be
gauge-independent). As mentioned above, our chosen vector field will be the
one associated with the azimuthal coordinate, namely $\left( \frac{\partial}
{\partial \varphi} \right)^{i}$.

Another important feature is the construction of a preferred coordinate
system, so that the Legendre polynomials associated with spherical harmonics
will possess the right orthonormality properties; as expressed in the
implementation of \cite{SchKriBey06}, this reduces to finding a set of
coordinates $(\zeta, \varphi)$ where the metric on the two-surface can be written as: 
\begin{equation}
\label{e:2dflat}
q^{\mathcal{H}}_{ab} = R_{\mathcal{H}}^{2}\left(
  f(\zeta)^{-1} D_{a}\zeta D_{b}\zeta + f(\zeta) D_{a}\varphi
  D_{b}\varphi \right),
\end{equation}
with $R_{\mathcal{H}}$ the areal radius of the sphere and $f(\zeta)$ determined in terms
of the two-dimensional Ricci scalar and the norm of $\varphi^a$ \cite{AshENPVDB}. In the
axisymmetric case studied here for the horizon, the integral curves for the
coordinate $\varphi$ are already defined by the orbits of the vector field $ (
\frac{\partial} {\partial \varphi})^{i}$. The coordinate $\zeta$ is defined by
\begin{equation}
\label{e:defzeta}
 D_{a}\zeta = \frac{1}{R_{\mathcal{H}}^{2}} \epsilon_{ba} \varphi^{b}.
\end{equation}

An appropriate normalization should be added, that ensures that
$\oint_{\mathcal{H}} \zeta d^{2}V =0$. In the Kerr case, those coordinates
turn out to correspond with the Boyer-Lindquist coordinates, with $\zeta =
\cos{\theta}$ in spherical coordinates \cite{SchKriBey06}.

The mass and angular momentum multipoles of order $n$ are then defined, by
analogy with electromagnetism \cite{AshENPVDB}:
\begin{eqnarray}
\label{e:multipoles}
M_{n} &=&  \frac{R_{\mathcal{H}}^{n}M_{\mathcal{H}}}{8 \pi}
\oint_{\mathcal{S}} \{ \mathcal{R} P_{n} (\zeta)\} d^{2}V,\\
J_{n} &=& \frac{R_{\mathcal{H}}^{n-1}}{8\pi}
\oint_{\mathcal{S}}P^{'}_{n}(\zeta) K_{ab}s^{a}\varphi^{b} d^{2}V.
\end{eqnarray}

\begin{figure}
\centering
\includegraphics*[width = 8cm]{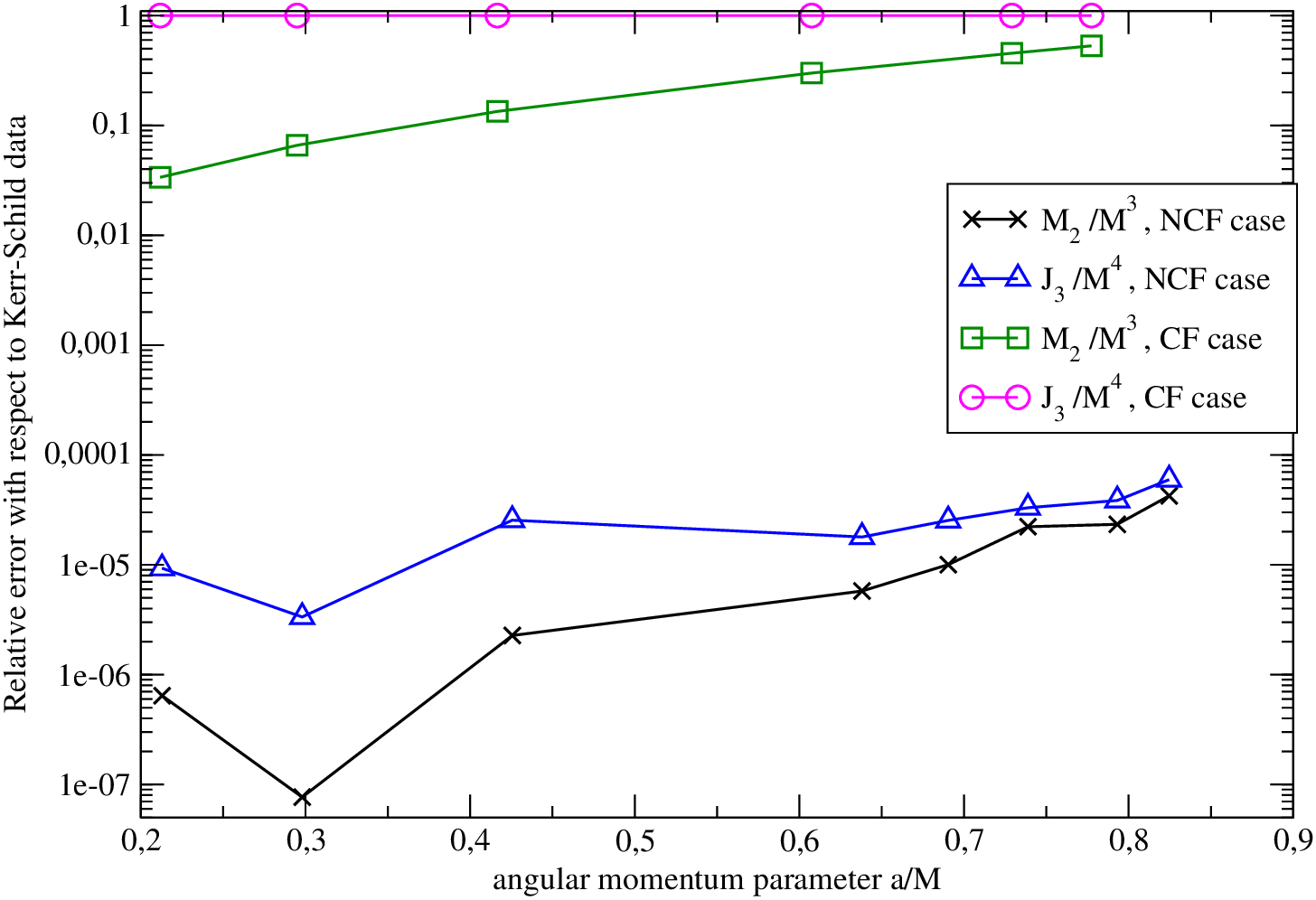}
\caption{Computations of the rescaled second-order mass multipole $\left(\frac
    {M_{2}}{M^3}\right)$ and the rescaled third-order angular multipole $\left(
    \frac {J_{3}} {M^{4}}\right)$. Relative differences with respect
    to values for a
  Kerr-Schild analytical horizon are displayed.}
\label{f:Multipoles}
\end{figure}

With this definition and using the Gauss-Bonnet theorem it is trivial to see
that $M_{0}=M_{\mathcal{H}}$ and $ J_{1}= J_{K}$, the Komar angular momentum
on the horizon.

We should emphasize that these multipoles, except for $M_{0}$ and $J_{1}$, are
in general different from the field gravitational multipoles that can be
defined at infinity. However, the authors in \cite{AshENPVDB} have
pointed out that the knowledge of all the multipoles of an isolated horizon allows to
reconstruct the whole horizon, and also the spacetime in a vicinity of this
horizon. The multipoles then discriminate exactly every isolated
horizon, and the spacetime at its vicinity. Figure~\ref{f:Multipoles} shows the
capacity of telling apart the horizon of a CF axisymmetric slice and the one
of a NCF slice, in Dirac gauge. Data are also compared with an analytic Kerr
solution in Kerr-Schild coordinates. Apart from the accuracy obtained for our
NCF data (and a further confirmation that we indeed have obtained the actual
Kerr spacetime), we see the clear distinction made by this computation between
the Kerr horizon and a conformal approximation of it. Together with Dain's number, this study has proven that those two tools are
very well-suited to study isolated horizon properties, and the
distance between data obtained from, say an evolution scheme, and the eventual
equilibrium black hole data it is supposed to reach. Ultimate tests on the
characterization of the obtained data as slices of Kerr could be
achieved by implementing the schemes proposed in \cite{GarVal08,FerSae08}.

%%%%%%%%%%%%%%%%%%%%%%%%%%%%%%%%%%%%%%%%%%%%%%%%%%%%%%%%%%%%%%%%%%%%%%%%%%%%%%%%%
\section {Discussion}\label{s:discus}
%%%%%%%%%%%%%%%%%%%%%%%%%%%%%%%%%%%%%%%%%%%%%%%%%%%%%%%%%%%%%%%%%%%%%%%%%%%%%%%%% 

The data we get with our simulations are interesting at several levels. They
allow to make a direct comparison between the conformally flat approximation and
the exact solution for axisymmetric spacetimes containing a black hole, that are both calculated {\it a
  priori\/} and in the same gauge. As we have seen in Sec.~\ref{s:resus}, 
this gives us insight about the geometric features of the exact solution; we
can single important issues concerning for example the intrinsic geometry of the
horizon, via multipoles and the Penrose inequalities. Numerical tools are in this
respect implemented and their efficiency tested.

At a more theoretical level, the method we used to get those data is a little
bit heterodox: providing standard non-expanding horizon conditions for (3+1) variables such as
$\beta^{i}$, $\psi$ and $N\psi$, we choose in addition not to prescribe any further
geometrical information for the conformal part, symbolized here by the
tensorial field $h^{ij}$. This has been motivated by the fact that, given the
tensorial equation corresponding to $h^{ij}$ in our formalism, it appears that
we most likely {\em cannot\/} prescribe anything else in the studied setting. By numerical
transformation of the operator acting on $h^{ij}$ we ensured that at every
iteration step no boundary condition was required. The fact that our system of
Einstein equations written this way converges to the required Kerr solution
shows that indeed, no additional information was needed in the single
horizon case. To be more precise,
our method suggests that in this case, the conformal geometry of the MOTS is
directly encoded in the Einstein equations, when written with an equilibrium
ansatz: we directly use these equations to justify a no-boundary treatment.

In the light of this numerical study, we can make a parallel with the
proposition made in \cite{CookBaum08}. In that paper, the authors suggested,
after a gauge dependence analysis for $\tilde{\gamma}^{ij}$ on the horizon,
that a prescription could be made for the conformal geometry on the horizon.
In this respect, they justify, for the projection of the three-metric on the
two-surface, the following choice: 
\begin{equation}
\label{e:propcook}
 q_{ab} = \omega^{2}f_{ab},
\end{equation}
with $f_{ab}$ the usual diagonal round metric for a two-sphere in spherical
coordinates adapted to the horizon. This choice should not affect the physics
of the three-slice, and suffices to recover the solution for Einstein
dynamical equation with a slice of a spacetime containing an isolated horizon. Their
study is made in a differential gauge generalizing the Dirac gauge we are
using, namely $\mathcal{D}_{i}\tilde{\gamma}^{ij} = V^{i}$, with $V^{i}$ a
regular vector field on the three-slice. Our case corresponds to $
V^{i} = 0$, which is precisely the one treated in detail in \cite{CookBaum08}.

When comparing to our data, we find that the projection of our three-metric on the
two-surface is not conformally related to the flat metric in adapted spherical
coordinates. This means that in our particular
case, and in regard of the particular no-boundary argument we use, a boundary
condition of the same type as~(\ref{e:propcook}) is probably inconsistent with
our data, and is likely a choice that we do not have the freedom to
make (note however that geometric conditions in
\cite{Jaram08} making full
use of the isolated horizon structure are indeed compatible with the 
present results, i.e. they are identically satisfied in the present Kerr case,
whose horizon is indeed an isolated horizon). 
Unfortunately, the authors in \cite{CookBaum08} did not present any
numerical results to support their claims, that we could have compared with
ours.

We insist here on an important caveat for our argumentation: assumptions can
only be justified in the very particular case we are studying here, which is
the axisymmetric vacuum spacetime. This spacetime has very specific and
non-trivial properties, all related to the uniqueness theorem of Carter
\cite{Carter72}. Although the reasoning we have made on Sec.~\ref{s:FCF}
for the operator could apply in other isolated horizon studies, we are not
certain that our algorithm would globally converge when applied to a more
general case (e.g. a black-hole binary system); a failure of this behavior
would probably mean that an additional information about the conformal
two-geometry has to be given to the system.
 Geometric fully isolated horizon boundary conditions proposed in \cite{Jaram08} could then
be enforced (note that geometric inner boundary conditions in \cite{Jaram08}
are not necessarily tied to the particular analytic setting here discussed and, more
generally, they would also apply in schemes not enforcing the coordinate
adaptation to stationarity at the horizon, $b=N_{\mathcal{H}}$, crucial for the 
singular nature of operators (\ref{e:opQ})).
 
Finally, let us point out the fact that we made those simulations by
prescribing only the geometry of the horizon, and the geometry of spacetime at
infinity. No assumption has been made for axisymmetry in the three-slice
(computations can easily be made in the full 3D case and give the same
results). Prescribing a vanishing expansion and a conformal Killing symmetry
on a horizon, together with asymptotically flat hypothesis, our code converges
to the only solution of the Kerr spacetime. Without claiming any rigorous
demonstration here, this numerical result is most likely a support to the well
known black hole rigidity theorem \cite{HawEll73}, where the same hypotheses
lead to a uniqueness theorem involving the Kerr solution as the only one with no
electromagnetic field.

%%%%%%%%%%%%%%%%%%%%%%%%%%%%%%%%%%%%%%%%%%%%%%%%%%%%%%%%%%%%%%%%%%%%%%%%%%%%%%%%%
\section{Conclusion}\label{s:conc}
%%%%%%%%%%%%%%%%%%%%%%%%%%%%%%%%%%%%%%%%%%%%%%%%%%%%%%%%%%%%%%%%%%%%%%%%%%%%%%%%%

We have used the prescription for a fully constrained scheme of (3+1) Einstein
equations in generalized Dirac gauge \cite{Bon04,Cord1} to retrieve stationary axisymmetric black
hole spacetime, and compared it with the analytical solution of Kerr type. An advanced handling of the conformal geometry of our three-slice allowed us
to reach actual stationarity with good resolution precision for our
scheme. Although we used standard quasi-equilibrium conditions concerning
boundary values for other metric fields in the excised horizon, we found that
the conformal geometry on the horizon required no prescription whatsoever in
the single horizon case. This is in contrast with suggestions available in the
literature \cite{CookBaum08}, and probably suggests an underlying physical feature of the
horizon geometry (maybe related to uniqueness of the Kerr solution). To our
knowledge, it is the first time the conformal part is numerically
computed in a black hole spacetime using only a prescription on the
stationarity of spacetime (and without resorting to additional symmetries). The application of this feature to the more general
initial data problem is evident: in the same spirit as the work done in
\cite{ShibLim} for neutron-star binaries, using it for the black-hole binary
system could lead to significant improvement in the available initial data for
evolution codes. Further numerical work will clarify this issue.

We have implemented and used in our study numerical tools aimed at
characterizing the geometry and physical properties related to horizons
embedded in spacetime; those tools, among which a complete multipole analysis
for two-surfaces as gravitational sources, have proven very accurate for
diagnostics involving the horizon geometry and physical features. They will be
more thoroughly presented, and tested in more general cases, in an upcoming
work.

%%%%%%%%%%%%%%%%%%%%%%%%%%%%%%%%%%%%%%%%%%%%%%%%%%%%%%%%%%%%%%%%%%%%%%%%%%%%%%%%%
\acknowledgments The authors warmly thank \'Eric Gourgoulhon and
Philippe Grandcl\'ement for many valuable
discussions, and a careful reading of the manuscript. We also thank the referee for numerous useful comments. 
NV and JN were supported
by the A.N.R. Grant BLAN07-1\_201699 entitled ``LISA Science''.
JLJ has been supported by the Spanish MICINN under project FIS2008-06078-C03-01/FIS and 
Junta de Andaluc\'{\i}a under projects FQM2288, FQM219.
%%%%%%%%%%%%%%%%%%%%%%%%%%%%%%%%%%%%%%%%%%%%%%%%%%%%%%%%%%%%%%%%%%%%%%%%%%%%%%%%%
%%%%%%%%%%%%%%%%%%%%%%%%%%%%%%%%%%%%%%%%%%%%%%%%%%%%%%%%%%%%%%%%%%%%%%%%%%%%%%%%%
%%%%%%%%%%%%%%%%%%%%%%%%%%%%%%%%%%%%%%%%%%%%%%%%%%%%%%%%%%%%%%%%%%%%%%%%%%%%%%%%%%%
%%%%%%%%%%%%%%%%%%%%%%%%%%%%%%%%%%%%%%%%%%%%%%%%%%%%%%%%%%%%%%%%%%%%%%%%%%%%%%%%%%%
%%%%%%%%%%%%%%%%%%%%%%%%%%%%%%%%%%%%%%%%%%%%%%%%%%%%%%%%%%%%%%%%%%%%%%%%%%%%%%%%%%%

\appendix

\section{Tensor spectral quantities adapted to the Dirac Gauge}
\label{s:appendix}

 We here give the definition of the three spectral quantities introduced
 in Sec.~\ref{ss:non_cf}, that describe the divergence-free degrees
 of freedom (with respect to the Dirac gauge) associated with a rank two symmetric
 tensor. The reader is
 also invited to go to \cite{Cord2} or \cite{NovCV} where more detailed
 calculations are provided.
  
 We first define a set of spin-weighted tensor spherical harmonics components for a
 symmetric rank-2 tensor, directly linked to the tensor spherical harmonics as
 introduced by Mathews and Zerilli~\cite{MathZer1,MathZer2}. We shall give the expression for
 these components of the tensor $h^{ij}$ using the classical
 spherical coordinate basis, which is used in practice in our
 computations. With the notation $P = h^{\th \th} + h^{\ph \ph}$, the six pure
 spherical harmonics components of $h^{ij}$ are defined as :
\begin{eqnarray}
  \lapang \eta &=& \frac{\partial h^{r\th}}{\partial \th} + \frac{h^{r\th}}{\tan
    \th} + \frac{1}{\sin \th} \frac{\partial h^{r\ph}}{\partial
    \ph},  \\ 
  \lapang \mu &=& \frac{\partial h^{r\ph}}{\partial \th} + \frac{h^{r\ph}}{\tan
    \th} - \frac{1}{\sin \th} \frac{\partial h^{r\th}}{\partial \ph},
   \\
  \lapang \left( \lapang + 2 \right) \mw &=& \frac{\partial^2 P}{\partial
    \th^2} + \frac{3}{\tan \th} \frac{\partial P}{\partial \th} -
  \frac{1}{\sin^2 \th} \frac{\partial^2 P}{\partial \ph^2} \nonumber \\
   -2P &+& \frac{2}{\sin
    \th} \frac{\partial}{\partial \ph} \left( \frac{\partial
      h^{\th\ph}}{\partial \th} +
    \frac{h^{\th\ph}}{\tan \th} \right) ,\\
  \lapang \left( \lapang + 2 \right) \mx &=& \frac{\partial^2
    h^{\th\ph}}{\partial \th^2} + \frac{3}{\tan \th} \frac{\partial
    h^{\th\ph}}{\partial \th} - \frac{1}{\sin^2 \th} \frac{\partial^2
    h^{\th\ph}}{\partial \ph^2} \nonumber \\
-2h^{\th\ph} &-& \frac{2}{\sin \th}
  \frac{\partial}{\partial \ph} \left( \frac{\partial P}{\partial \th} +
    \frac{P}{\tan \th} \right),
 \end{eqnarray}
the fifth and sixth scalar fields being simply the tensor trace $h$
with respect to the flat metric and the
$h^{rr}$ spherical component. Let us note that these relations are
more tractable when using
a scalar spherical harmonics decomposition (introduced in
Sec.~\ref{ss:BChij}) for all fields. Indeed, an angular
Laplace operator acting on a field reduces then to a simple algebraic
operation on every spherical harmonics component. 
Inverse relations can also be computed to retrieve the classical components
of $h^{ij}$ from spherical harmonics quantities.

 We now derive the main variables related to our study: with the divergence-free
decomposition   $h^{ij} = \mathcal{D}^{i}W^{j} +  \mathcal{D}^{j}W^{i} + h^{ij}_{T}$,
and $\mathcal{D}_{i} h^{ij}_{T} = 0$, a choice for three quantities
defined from $h^{ij}$ and
verifying:
 \begin{equation}
 h^{ij}_{T}=0 \Rightarrow A = B = C = 0,
\label{e:def2abc}
\end{equation}
can be expressed as the following scalar fields (see \cite{NovCV}):
\begin{eqnarray}
  A = \der{{\cal X}}{r} - \frac{\mu}{r}, \label{e:def_A}\\
  B = \der{{\cal W}}{r} - \frac{\lapang {\cal W}}{2r} -
  \frac{\eta}{r} + \frac{h - h^{rr}}{4r},
  \label{e:def_B}\\
  C = \der{(h - h^{rr})}{r} - \frac{3h^{rr}}{r} + \frac{h}{r} - 2\lapang \left( \der{{\cal W}}{r} +
    \frac{{\cal W}}{r} \right) \label{e:def_C}
\end{eqnarray}

 These quantities can also be decomposed onto a scalar spherical harmonics basis.
 The equivalence in~(\ref{e:def2abc}) is achieved up to boundary
 conditions.

 To show how the quantities $A$, $B$ and $C$ behave with respect to the
 Laplace operator, we shall assume in the following that the tensor $h^{ij}$
 is the solution of a Poisson equation of the type $\Delta h^{ij} = S^{ij}$.
 We can deduce a scalar elliptic system verified by $A$, $B$ and $C$ as:
\begin{eqnarray}
 \Delta A = A_{S} \label{e:deltaa}\\
 \Delta B - \frac{C}{2r^{2}} = B_{S} \label{e:deltab}\\
  \Delta C + \frac{2C}{r^{2}} + \frac{8\lapang B}{r^{2}} = C_{S} \label{e:deltac},
\end{eqnarray}
 Where $A_{S}$, $B_{S}$ and $C_{S}$ are the corresponding quantities
 associated with the source $S^{ij}$. 
 A simple way of decoupling the last two elliptic equations is to
 define the variables  $\tilde{B} = \sum_{l, m} \tilde{B}^{l
   m} Y_{l m}$ and  $\tilde{C} = \sum_{l, m} \tilde{C}^{l m} Y_{l m}$
  with:
\begin{eqnarray}
    \tilde{B}^{lm} = B^{l m} + \frac{C^{lm}}{2(l+1)},\\
    \tilde{C}^{l m} = C^{l m} -4l B^{lm}. \label{e:deftilde}
\end{eqnarray}
   Thus, we can write an equivalent system for~(\ref{e:deltaa},\ref{e:deltab},\ref{e:deltac}) as:
\begin{eqnarray}
\Delta A &=& A_{S} \label{e:deltaa2}, \\
\tilde{\Delta} \tilde{B} &=& \tilde{B}_{S} \label{e:deltatb}, \\
\Delta^{*} \tilde{C} &=& \tilde{C}_{S} \label{e:deltatc}, 
\end{eqnarray}
With the following elliptic operators defined for each spherical harmonic index l:

\begin{eqnarray}
\Delta &=& \frac{\partial^{2}}{\partial r^{2}} + \frac{2}{r}
\frac{\partial}{\partial r} -\frac{l (l+1)}{r^{2}}I\\
\tilde{\Delta} &=& \Delta + \frac{2l}{r^{2}}I, \\
\Delta^{*} &=& \Delta - \frac{2(l+1)}{r^{2}}I.
\end{eqnarray}

$I$ is the identity operator. $A$, $\tilde{B}$ and $\tilde{C}$ are then defined as three scalar fields 
characterizing only the divergence-free part of a symmetric rank two tensor $h^{ij}$, and giving a system of three
decoupled scalar elliptic equations in the Poisson problem for this tensor. Hence they are very well suited to the
study of a tensorial elliptic problem in Dirac gauge. 
Let us finally note that the quantities $\tilde{B}$ and $\tilde{C}$
are directly related to each other by the trace of the considered
tensor \cite{NovCV}:
If we know {\em a priori} the value of the trace for $h^{ij}$, then the knowledge of $\tilde{B}$ suffices to recover $\tilde{C}$ with no 
additional information (the converse being equally true).

\section{Recovery of $h^{ij}$ from $A$ and $\tilde{B}$}
\label{s:appendixB}
In this section, we come up with technical details for resolution of the gauge
differential system introduced in Sec.~\ref{ss:non_cf}, to reconstruct the
tensor $h^{ij}$ from the quantities $A$ and $\tilde{B}$.

We begin by expressing components of the vector field representing the
divergence of $h^{ij}$:
\begin{equation}
H^{i} = \mathcal{D}_{i}h^{ij},
\end{equation}
with the Dirac gauge for $H^i=0$. Adopting the vector spherical harmonics
decomposition suggested in \cite{Bon04}, the three spherical harmonics
components of $H^{i}$ are expressed, in function of the spherical harmonics
components of $h^{ij}$ (see Appendix~\ref{s:appendix}), as:
\begin{eqnarray}
  H^r &=& \frac{\partial h^{rr}}{\partial r} + \frac{3h^{rr}}{r} + \frac{1}{r}
  \left( \lapang \eta - h \right), \label{e:Hr}\\
  H^\eta \! &=& \! \lapang \left[ \frac{\partial \eta}{\partial r} +
    \frac{3\eta}{r} + \frac{1}{r} \left( \left(\lapang + 2 \right)\mw +
      \frac{h - h^{rr}}{2} \right) \right], \label{e:Heta}\\
  H^\mu &=& \lapang \left[ \frac{\partial \mu}{\partial r} +
    \frac{3\mu}{r} + \frac{1}{r} \left( \lapang + 2 \right) \mx \right]. \label{e:Hmu}
\end{eqnarray}

Those three expressions, alongside with definitions of the quantities
$A$ and $\tilde{B}$, will allow to express two decoupled
differential systems. The first one, involving the spherical harmonics
components $\mu$ and $\mathcal{X}$, combines the expression for the
scalar field $A$, as well as the fact that $H^{\mu}$ vanishes under
the Dirac gauge :
\begin{equation}
 \left\{
\begin{aligned}
  &\der{{\cal X}}{r} - \frac{\mu}{r} = A, \\
   & \frac{\partial \mu}{\partial r} +
    \frac{3\mu}{r} + \frac{1}{r} \left( \lapang + 2 \right) \mx
     = 0.
\end{aligned}
\right. 
\label{e:diracmag}
\end{equation}

The second system is composed of the definition of $\tilde{B}$ for
each of its spherical harmonic component $\tilde{B}^{lm}$, as well
as the vanishing of $H^{r}$ and $H^{\eta}$(again, due to Dirac gauge):
\begin{equation}
 \left\{
\begin{aligned}
  &\tilde{B}^{lm} = B^{l m} + \frac{C^{lm}}{2(l+1)},\\
 &\frac{\partial h^{rr}}{\partial r} + \frac{3h^{rr}}{r} + \frac{1}{r}
  \left( \lapang \eta - h \right) =0, \\
 &\frac{\partial \eta}{\partial r} +
    \frac{3\eta}{r} + \frac{1}{r} \left( \left(\lapang + 2 \right)\mw +
      \frac{h - h^{rr}}{2} \right) =0.
\end{aligned}
\right. 
\label{e:diracel}
\end{equation}
with the expressions (\ref{e:def_B}, \ref{e:def_C}) of $B$ and $C$ as
functions of the spherical harmonics components of $h^{ij}$. In this system,
the trace is given {\it a priori}, so that only three spherical harmonics components
are considered as unknowns.

We refer to the analysis of \cite{NovCV} to affirm that, when solving our
equations in $\mathbb{R}^{3}$ minus an excised inner sphere, one boundary
condition has to be provided at the surface for the system (\ref{e:diracmag}),
and two for the system (\ref{e:diracel}). As pointed out in
Sec.\ref{ss:BChij}, these conditions are retrieved as compatibility conditions
based on the original elliptic tensorial equation. Overall, we are able to
invert the two Dirac differential systems, and retrieve all the spherical
harmonics components of $h^{ij}$ from the sole knowledge of $A$, $\tilde{B}$
and the trace $h$.

%%%%%%%%%%%%%%%%%%%%%%%%%%%%%%%%%%%%%%%%%%%%%%%%%%%%%%%%%%%%%%%%%%%%%%%%%%%%%%%%
%%%%%%%%%%%%%%%%%%%%%%%%%%%%%%%%%%%%%%%%%%%%%%%%%%%%%%%%%%%%%%%%%%%%%%%%%%%%%%%%%
%%%%%%%%%%%%%%%%%%%%%%%%%%%%%%%%%%%%%%%%%%%%%%%%%%%%%%%%%%%%%%%%%%%%%%%%%%%%%%%%%%%
%%%%%%%%%%%%%%%%%%%%%%%%%%%%%%%%%%%%%%%%%%%%%%%%%%%%%%%%%%%%%%%%%%%%%%%%%%%%%%%%%%%
%%%%%%%%%%%%%%%%%%%%%%%%%%%%%%%%%%%%%%%%%%%%%%%%%%%%%%%%%%%%%%%%%%%%%%%%%%%%%%%%%%%

\end{document}